\documentclass[11pt]{article}
\pdfoutput=1
\usepackage{jheppub}
\usepackage{graphicx}
\usepackage{amssymb}
\usepackage{amsmath}
\usepackage{amsfonts}
\usepackage{latexsym}
\usepackage{hyperref}
\usepackage{physics}
\usepackage{bbm}
\usepackage{empheq}
\usepackage{tikz}
\usepackage{cancel}
\usepackage{hhline}
\usepackage{caption}
\usepackage{soul}
\usepackage{ulem}

\usetikzlibrary{arrows.meta}
\usetikzlibrary{decorations.markings}
\tikzset{->-/.style={decoration={
			markings,
			mark=at position #1 with {\arrow{>}}},postaction={decorate}}}

\newcommand\nn{\nonumber}
\newcommand\fft[2]{\frac{#1}{#2}}

\newcommand\mI{\mathcal{I}}
\newcommand\tu{\tilde{u}}

\newcommand\mW{\mathcal{W}}
\newcommand\tn{\tilde{n}}
\newcommand\tDelta{\tilde{\Delta}}

\newcommand\tq{\tilde{q}}

\newcommand\mF{\mathcal{F}}

\newcommand\mn{\mathfrak{n}}
\newcommand\tmn{\tilde{\mathfrak{n}}}
\newcommand\mm{\mathfrak{m}}

\newcommand\mft{\mathfrak{t}}

\newcommand\mV{\mathcal{V}}

\newcommand\bs{\bar{s}}
\newcommand\bU{\bar{U}}
\newcommand\mN{\mathcal{N}}
\def\ri{{\rm i}}
\newcommand\mR{\mathcal{R}}
\newcommand\mQ{\mathcal{Q}}

\newcommand\mO{\mathcal{O}}

\newcommand\mg{\mathfrak{g}}

\newcommand\mZ{\mathcal{Z}}

\makeatletter
\newcommand*{\rom}[1]{\expandafter\@slowromancap\romannumeral #1@}
\makeatother

\begin{document}
	
\preprint{KIAS-P24009}
\title{Superconformal Indices of 3d $\mathcal N=2$ SCFTs and Holography}
	
\author[a]{Nikolay Bobev,}
\author[b]{Sunjin Choi,}
\author[c]{Junho Hong,}
\author[d]{and Valentin Reys}

\affiliation[a]{Institute for Theoretical Physics, KU Leuven\,,\\ Celestijnenlaan 200D, B-3001 Leuven, Belgium}
\affiliation[b]{School of Physics, Korea Institute for Advanced Study\,,\\ 85 Hoegi-ro, Dongdaemun-gu, Seoul 02455, Republic of Korea}
\affiliation[c]{Department of Physics \& Center for Quantum Spacetime, Sogang University\,,\\ 35 Baekbeom-ro, Mapo-gu, Seoul 04107, Republic of Korea}
\affiliation[d]{Université Paris-Saclay, CNRS, CEA, \\
	Institut de physique théorique, 91191, Gif-sur-Yvette, France}

\emailAdd{nikolay.bobev@kuleuven.be}
\emailAdd{sunjinchoi@kias.re.kr}
\emailAdd{junhohong@sogang.ac.kr}
\emailAdd{valentin.reys@ipht.fr}

%%%%%
	
\abstract{We study the superconformal index of 3d $\mN=2$ superconformal field theories on $S^1\times_{\omega} S^2$ in the Cardy-like limit where the radius of the $S^1$ is much smaller than that of the $S^2$. We show that the first two leading terms in this Cardy-like expansion are dictated by the Bethe Ansatz formulation of the topologically twisted index of the same theory. We apply this relation to 3d $\mN=2$ holographic superconformal field theories describing the low-energy dynamics of $N$ M2-branes and derive closed form expressions, valid to all orders in the $1/N$ expansion, for the two leading terms in the Cardy-like expansion of the superconformal index. We also discuss the implications of our results for the entropy of supersymmetric Kerr-Newman black holes in AdS$_4$ and the four-derivative corrections to 4d gauged supergravity.}
	
\maketitle \flushbottom

%%%%%
\section{Introduction}\label{sec:intro}
%%%%%

The interplay between supersymmetric localization and holography has led to valuable precision test of the AdS/CFT correspondence and a new vantage point towards the structure of string and M-theory on non-trivial flux backgrounds, see \cite{Pestun:2016zxk,Zaffaroni:2019dhb} for reviews and further references. A prominent role in these holographic explorations is played by the superconformal indices, or $S^1\times_{\omega}S^{d-1}$ supersymmetric partition functions. They capture the degeneracy of supersymmetric states in the SCFT and provide a microscopic counting for the entropy of the dual supersymmetric Kerr-Newman black holes in AdS$_{d+1}$. The goal of this work is to continue the exploration of the relation between supersymmetric partition functions and holography in the context of the superconformal index (SCI) of 3d $\mathcal{N}=2$ SCFTs \cite{Bhattacharya:2008bja,Bhattacharya:2008zy,Kim:2009wb}.

The SCI can be viewed as a supersymmetric partition function $\mathcal{I}$ on the Euclidean manifold $S^1\times_{\omega} S^2$ where $\omega$ is the fugacity for the angular momentum on $S^2$. Using supersymmetric localization, one can derive a matrix integral expression for the SCI that facilitates its evaluation for a large class of $\mN=2$ supersymmetric field theories, not necessarily requiring conformal symmetry \cite{Imamura:2011su,Imamura:2011uj,Krattenthaler:2011da,Kapustin:2011jm}. The resulting matrix integral is a complicated function of the various parameters of the SCFT, like real masses or flavor symmetry fugacities, which is hard to evaluate for general values of $\omega$. One way to make progress is to focus on the Cardy-like limit of the SCI where the radius of the $S^1$ is taken to be much smaller than the one of the $S^2$, see \cite{Choi:2019zpz,Choi:2019dfu,Nian:2019pxj}. Due to supersymmetry, this geometric limit amounts to taking the angular fugacity $\omega$ to vanish. As shown in \cite{Choi:2019dfu, Choi:2019zpz}, in the $\omega \to 0$ limit of the SCI, the two leading terms are of order $\omega^{-1}$ and $\omega^0$.\footnote{The Cardy-like expansion in $1/ \omega$ of the SCI of 4d $\mathcal{N}=1$ SCFTs terminates at order $\omega^1$, see \cite{Cassani:2021fyv,ArabiArdehali:2021nsx,GonzalezLezcano:2020yeb,Ardehali:2021irq}. As we show explicitly in Section~\ref{sec:general}, this is not the case for 3d $\mathcal{N}=2$ theories, where one generically finds an infinite series in powers of $\omega$.} Moreover, it was noted in \cite{Choi:2019dfu, Choi:2019zpz} that in some theories, these two leading terms have an intimate relation to another supersymmetric partition function known as the topologically twisted index (TTI). The TTI is a partition function of 3d $\mathcal{N}=2$ SCFTs placed on $S^1\times \Sigma_{\mathfrak{g}}$ with a partial topological twist on the Riemann surface of genus $\mathfrak{g}$ by the superconformal ${\rm U}(1)$ $R$-symmetry of the theory \cite{Benini:2015noa,Benini:2016hjo,Closset:2016arn,Closset:2017zgf}. Using supersymmetric localization, the TTI can be reduced to a matrix integral, which in turn can be evaluated by a version of the residue theorem and recast as a system of algebraic equations. An important role in this ``Bethe Ansatz'' approach to the evaluation of the TTI is played by the so-called Bethe potential $\mathcal{V}$.  The observation of \cite{Choi:2019dfu, Choi:2019zpz}, further studied in \cite{Bobev:2022wem}, is that the $\omega^{-1}$ term in the Cardy-like expansion of the SCI is determined by the Bethe potential, while the $\omega^0$ term is fixed by the logarithm of the TTI on $S^1\times S^2$. 

Recently, in \cite{Bobev:2022wem}, we studied this relation between the SCI and the TTI in the context of two models arising from $N$ M2-branes in M-theory, namely the ABJM theory \cite{Aharony:2008ug} and the 3d $\mathcal{N}=4$ ADHM theory (also called the $N_f$ matrix model) \cite{Mezei:2013gqa,Grassi:2014vwa,Porrati:1996xi}. It was shown in \cite{Bobev:2022jte,Bobev:2022eus,Bobev:2023lkx} that the Bethe potential and TTI for these two models can be computed to all orders in the $1/N$ expansion using precise numerical techniques that lead to exact analytic expressions. These results were then employed in \cite{Bobev:2022wem} to derive the first two leading terms in the Cardy-like expansion of the SCI for these theories. The explicit expressions for the SCI for these two models can then be expanded at large $N$ and the leading $N^{\frac{3}{2}}$, $N^{\frac{1}{2}}$ and $\log N$ terms can be matched to the on-shell action of the dual supersymmetric Euclidean Kerr-Newman black hole solution embedded in M-theory, see \cite{Choi:2019zpz,Choi:2019dfu,Nian:2019pxj,Bobev:2022wem,Bobev:2023dwx}.

The goal of this work is to extend and generalize these results in several ways. In Section~\ref{sec:general}, we build on the results of \cite{Choi:2019dfu, Choi:2019zpz,Bobev:2022wem} to provide a general derivation of the relation between the order $\omega^{-1}$ and $\omega^{0}$ terms in the Cardy-like expansion of the SCI and the Bethe potential and the TTI. Importantly in this derivation, we do not assume that the 3d $\mathcal{N}=2$ theory at hand is holographic or has any kind of large $N$ limit. In Section~\ref{sec:ex}, we apply this relation between the SCI and the TTI to study the SCI for various 3d $\mathcal{N}=2$ SCFTs arising on the worldvolume of M2-branes. These theories have a dual holographic description in terms of M-theory on an asymptotically AdS$_4\times Y_7$ background, where the details of the SCFT are encoded in the geometry of the 7d Sasaki-Einstein (SE) manifold $Y_7$. For this class of models, we can leverage the numerical techniques for the calculation of the TTI discussed in \cite{Bobev:2022jte,Bobev:2022eus,Bobev:2023lkx} to find explicit expressions, valid to all orders in the $1/N$ expansion, for the first two leading terms in the Cardy-like expansion of the SCI. In Section~\ref{sec:holo}, we study the holographic implications of these field theory results for the SCI. We first formulate a prediction for the path integral of M-theory on the background of a supersymmetric Kerr-Newman black hole asymptotic to AdS$_4\times Y_7$ to all order in the $1/N$ expansion. We also show how the $N^{\frac{1}{2}}$ terms in the large $N$ expansion of the SCI and TTI can uniquely determine the four-derivative couplings in 4d $\mN=2$ minimal gauged supergravity following the approach of \cite{Bobev:2020egg,Bobev:2021oku}. This interplay between field theory and supergravity allows for a number of consistency checks of our results and leads to new predictions for the path integrals of the 3d SCFTs in question on other compact Euclidean manifolds. We also show that our results for the $\log N$ term in the large $N$ expansion of the SCI are in agreement with the recent calculations of logarithmic corrections to the entropy of supersymmetric Kerr-Newman black holes in AdS$_4$ \cite{Bobev:2023dwx}. We conclude our discussion with a summary of some open problems and possible generalizations of our work in Section~\ref{sec:discussion}. The five appendices contain our conventions for some of the special functions we use as well as various technical aspects of our calculations.

%%%%%
\section{Indices of 3d $\mN=2$ SCFTs}\label{sec:general}
%%%%%
In this section, we first briefly review the $S^1\times_{\omega} S^2$ superconformal index (SCI) and the $S^1\times S^2$ topologically twisted index (TTI) of 3d $\mN=2$ SCFTs. Next, we explain how the Cardy-like limit of the former is related to the latter, generalizing the observation of \cite{Bobev:2022wem} for the ABJM/ADHM theories to a larger class of $\mN=2$ SCFTs. We first summarize the main results schematically and then provide more details in the subsequent discussion.

%%%%%
\subsection{Summary of main results}\label{sec:general:summary}
%%%%%
Consider 3d $\mN=2$ Chern-Simons-matter quiver gauge theories with $p\in\mathbb{N}$ nodes, each of which represents a gauge group U$(N)_{k_r}$ with Chern-Simons (CS) level $k_r~(r=1,\ldots,p)$. The matter content of the quiver gauge theory consists of $\mathcal N=2$ chiral multiplets, collectively denoted by $\Psi$, in the $\mR_\Psi$ representation of the gauge group $G=\otimes_{r=1}^p$U$(N)_{k_r}$ with $R$-charge $R(\Psi)$ and flavor charges $f_x(\Psi)$, where $x$ runs over the dimension of the Cartan subgroup of the flavor group. Note that when $p=0$ the theory has no gauge group and consists only of interacting chiral multiplets. 

Our focus here is on the $\mN=2$ SCFTs living at the superconformal IR fixed point of the RG flow originating from the class of $\mN=2$ asymptotically free CS-matter quiver gauge theories described above in the UV.\footnote{Later we will also consider deformations of the SCFTs that break the conformal symmetry, and explore the relation between indices after such deformations. In a slight abuse of notation, we will still refer to these as $\mN=2$ SCFTs.} For these $\mN=2$ SCFTs, one can compute various important physical quantities protected by supersymmetry along the RG flow by applying supersymmetric localization to the UV CS-matter theories with explicit Lagrangians. Of particular interest in this work are the SCI and the TTI, and the relation between these two partition functions. \\

For the class of 3d $\mN=2$ SCFTs described above we show that the SCI in the Cardy-like limit $\omega\to \ri 0^+$ and the TTI are related in the following way, 
\begin{equation}
	\log[\text{SCI}]=-\fft{1}{\pi\omega}\Im[\text{Bethe potential}]+\Re[\log[\text{TTI}]]+\ri\varphi+\mO(\omega)\,.\label{summary}
\end{equation}
Here $q=e^{\ri\pi\omega}$ is the fugacity for the angular momentum on $S^2$, the Bethe potential is introduced in the Bethe-Ansatz (BA) formalism for the TTI discussed in Section~\ref{sec:general:TTI}, and the pure imaginary term $\ri\varphi$ arises from various phase factors that do not depend on the details of a given theory. We refer to the relation above as the \emph{Bethe formulation} of the SCI. The reason is that all quantities appearing on the r.h.s. of \eqref{summary} arise naturally in the Bethe formulation of the TTI described in \cite{Benini:2015noa,Benini:2015eyy}. It is worth emphasizing that the relation in (\ref{summary}) does not require any large $N$ limit or the existence of a holographic dual description of the CS-matter theory at hand. Put differently, it is a relation between two field theory partition functions, which is non-perturbative in $N$ and holds for all $\mN=2$ SCFTs described above. \\

For illustration, here we present perhaps the simplest example that exhibits the relation (\ref{summary}). Consider an 3d $\mN=2$ theory with two chiral multiplets and a superpotential
\begin{equation}
	W=\Psi_1(\Psi_2)^2\,.
\end{equation}
Each chiral multiplet $\Psi_i$ is charged under a U(1)$_R$ $R$-symmetry and a U(1)$_F$ flavor symmetry with charges $(r_i,f_i)$ respectively. Since the superpotential has $R$-charge two and is neutral under the flavor symmetry the charges of the chiral multiplets should satisfy the constraints
\begin{equation}\label{eq:suppotcharges}
	r_1+2r_2=2\qquad\text{and}\qquad f_1+2f_2=0\,.
\end{equation}
The SCI of this theory can be computed by supersymmetric localization \cite{Imamura:2011su,Imamura:2011uj,Krattenthaler:2011da,Kapustin:2011jm} and reads
\begin{equation}
	\mI_W(q,\xi)=\mI_{\Psi_1}(q,\xi)\mI_{\Psi_2}(q,\xi)\,,\qquad\text{where}\qquad\mI_{\Psi_i}(q,\xi)=\fft{(\xi^{-f_i}q^{2-r_i};q^2)_\infty}{(\xi^{f_i}q^{r_i};q^2)_\infty}\,,\label{SCI:single}
\end{equation}
in terms of $\infty$-Pochhammer symbols for the fugacities $q = e^{\mathrm{i}\pi\omega}$ and $\xi$ associated with the U(1)$_R$ and U(1)$_F$ global symmetries. In the Cardy-like limit $\omega \rightarrow \ri0^+$, one can expand the single chiral multiplet contribution as
\begin{equation}
	\log\mI_{\Psi_i}(q,\xi)=-\fft{1}{\pi\omega}\Im\text{Li}_{2}(y_i)+\Re\log\bigg(\fft{y_i^{1/2}}{1-y_i}\bigg)^{1-\mn_i}+\mO(\omega)\,,\label{SCI:single:Cardy}
\end{equation}
where we have used the asymptotic expansion of the $\infty$-Pochhammer symbol (\ref{poch:asymp}) and the new parameters defined as
\begin{equation}
	y_iq^{\mn_i}=q^{r_i}\xi^{f_i}\,,\quad \text{with} \quad |y_i|=1\,,~\mathfrak{n}_i \in \mathbb{R}\,,~q\in\mathbb{R}\,.\label{SCI:redefine:single}
\end{equation}
The logarithm of the SCI, $\log\mI_W(q,\xi)$, then takes precisely the form of (\ref{summary}) provided the Bethe potential and the TTI for a single chiral multiplet are identified as
\begin{equation}
	\mV_{\Psi_i}=\text{Li}_{2}(y_i)\,,\qquad\text{and}\qquad Z_{\Psi_i}=\bigg(\fft{y_i^{1/2}}{1-y_i}\bigg)^{1-\mn_i}\,,\label{VTTI:single:1}
\end{equation}
respectively. Indeed, as we review in Section~\ref{sec:general:TTI}, the Bethe potential and the TTI for a single chiral multiplet are given by (\ref{VTTI:single:1}).

%%%%%
\subsection{Superconformal index}\label{sec:general:SCI}
%%%%%
The SCI, or $S^1\times_{\omega} S^2$ partition function, was defined in \cite{Bhattacharya:2008zy,Bhattacharya:2008bja,Kim:2009wb} and then analyzed for generic Lagrangian 3d $\mN=2$ SCFTs via supersymmetric localization in \cite{Imamura:2011su,Imamura:2011uj,Krattenthaler:2011da,Kapustin:2011jm,Aharony:2013dha,Aharony:2013kma}. It can be written as a trace over the Hilbert space of the theory in radial quantization, 
\begin{equation}
	\mI(q,\boldsymbol{\xi})=\Tr\relax\bigg[(-1)^{2j_3}e^{-\beta_1\{\mQ,\mQ^\dagger\}}q^{\Delta+j_3}\prod_x\xi_x^{F_x}\bigg]\, , \qquad\{\mQ,\mQ^\dagger\}=\Delta-R-j_3\,,\label{SCI:tr}
\end{equation}
where $\mQ$ is a supercharge, $\Delta$ is the energy in radial quantization, $R$ is the $R$-charge, $j_3$ is the third component of the angular momentum on $S^2$, and $F_x$ are charges associated with flavor symmetries. We introduce the chemical potential $\omega$ associated with the fugacity $q$ as $q=e^{\ri\pi\omega}$ and refer the reader to \cite{Imamura:2011su,Kim:2009wb,Bobev:2022wem} for its precise geometric meaning. Note that by the usual pairing argument, the index is independent of the circumference $\beta_{1}$ of the $S^1$ part of the geometry, see \cite{Bobev:2022wem} for a summary of the geometric interplay between $\beta_1$ and $\omega$. We note that the fermion number operator $(-1)^F$ in the  trace formula for the SCI of \cite{Imamura:2011su,Imamura:2011uj,Krattenthaler:2011da,Kapustin:2011jm} is replaced with $(-1)^{2j_3}$ in (\ref{SCI:tr}), which takes into account non-trivial phase contributions from magnetic monopoles based on the prescription of \cite{Aharony:2013dha,Aharony:2013kma}, see also \cite{Dimofte:2011py} for a related discussion.

The matrix integral expression of the SCI trace formula (\ref{SCI:tr}) for the class of 3d $\mathcal N=2$ SCFTs described in the previous subsection can be computed by supersymmetric localization and reads \cite{Imamura:2011su,Imamura:2011uj,Krattenthaler:2011da,Kapustin:2011jm,Aharony:2013dha,Aharony:2013kma}
\begin{equation}
	\begin{split}
		\mI(q,\boldsymbol{\xi})&=\fft{1}{(N!)^p}\sum_{\mm_1,\ldots,\mm_p\in\mathbb Z^N}\oint\bigg(\prod_{r=1}^p\prod_{i=1}^N\fft{dz_{r,i}}{2\pi \ri z_{r,i}}z_{r,i}^{k_r\mm_{r,i}}\xi_{T_r}^{\mm_{r,i}}\bigg)\\
		&\quad\times\prod_{r=1}^p\prod_{i\neq j}^Nq^{-\fft12|\mm_{r,i}-\mm_{r,j}|}\left(1-z_{r,i}z_{r,j}^{-1}q^{|\mm_{r,i}-\mm_{r,j}|}\right)\\
		&\quad\times\prod_{\Psi}\prod_{\rho_\Psi}(-1)^{\fft12(\rho_\Psi(\mm)+|\rho_\Psi(\mm)|)}\left(q^{1-R(\Psi)}e^{-\ri\rho_\Psi(h)}\prod_x\xi_x^{-f_x(\Psi)}\right)^{\fft12|\rho_\Psi(\mm)|}\\
		&\kern5em\times\fft{(e^{-\ri\rho_\Psi(h)}\prod_x\xi_x^{-f_x(\Psi)}q^{2-R(\Psi)+|\rho_\Psi(\mm)|};q^2)_\infty}{(e^{\ri\rho_\Psi(h)}\prod_x\xi_x^{f_x(\Psi)}q^{R(\Psi)+|\rho_\Psi(\mm)|};q^2)_\infty}\,,
	\end{split}\label{SCI:1}
\end{equation}
where we have turned on mixed CS terms between gauge and topological symmetries with the corresponding fugacities $\xi_{T_r}$. In the matrix model (\ref{SCI:1}), contour integrals for gauge zero modes $z_{r,i}=e^{\ri h_{r,i}}$ are over the unit circle, $\rho_\Psi$ runs over the weights of the representation $\mR_\Psi$ of the $\mathcal N=2$ chiral multiplet $\Psi$ with respect to the gauge group $G=\otimes_{r=1}^p\text{U}(N)_{k_r}$, and $\mm_{r,i}$ stand for integer-quantized gauge magnetic fluxes. We omit $N$ and $\{k_r\}$ in the argument of the SCI for notational convenience throughout this paper.

It is worth mentioning that the fugacities associated with the topological symmetries in the localization formula (\ref{SCI:1}) are not precisely the same as the ones introduced through the $F_x$ in the trace formula (\ref{SCI:tr}). We redefined them by absorbing extra phases in the localization formula arising from the replacement $(-1)^F\to(-1)^{2j_3}$ in the trace formula \cite{Aharony:2013dha,Aharony:2013kma}. In Appendix \ref{app:SCI-phase} we present the details of the derivation of the localization formula (\ref{SCI:1}) starting from the convention of \cite{Aharony:2013dha,Aharony:2013kma}.

%%%%%
\subsubsection{Factorization}\label{sec:general:SCI:fact}
%%%%% 
Evaluating the integrals over gauge zero modes and the sum over gauge magnetic fluxes in the localization formula (\ref{SCI:1}) is highly involved in general. In order to make progress, we will work in the Cardy-like limit $\omega \rightarrow \mathrm{i}0^+$ following \cite{Choi:2019dfu, Choi:2019zpz}. Firstly, it is useful to rewrite the localization formula (\ref{SCI:1}) as follows:
\begin{equation}
	\begin{split}
		\mI(\omega,\Delta,\mn)&=\fft{1}{(N!)^p}\sum_{\mm_1,\cdots,\mm_p\in\mathbb Z^N}\oint\bigg(\prod_{r=1}^p\prod_{i=1}^N\fft{dz_{r,i}}{2\pi \ri z_{r,i}}z_{r,i}^{k_r\mm_{r,i}}(y_{T_r}q^{-\mft_r})^{\mm_{r,i}}\bigg)\\
		&\quad\times\prod_{r=1}^p\prod_{i\neq j}^N(qz_{r,i}z_{r,j}^{-1})^{-\fft12|\mm_{r,i}-\mm_{r,j}|}\fft{(z_{r,i}^{-1}z_{r,j}q^{|\mm_i-\mm_j|};q^2)_\infty}{(z_{r,i}z_{r,j}^{-1}q^{2+|\mm_i-\mm_j|};q^2)_\infty}\\
		&\quad\times\prod_{\Psi}\prod_{\rho_\Psi}(-1)^{\fft12(\rho_\Psi(\mm)+|\rho_\Psi(\mm)|)}\left(q^{1-\mn_\Psi}e^{-\ri\rho_\Psi(h)}y_\Psi^{-1}\right)^{\fft12|\rho_\Psi(\mm)|}\\
		&\kern5em\times\fft{(e^{-\ri\rho_\Psi(h)}y_{\Psi}^{-1}q^{2-\mn_\Psi+|\rho_\Psi(\mm)|};q^2)_\infty}{(e^{\ri\rho_\Psi(h)} y_\Psi q^{\mn_\Psi+|\rho_\Psi(\mm)|};q^2)_\infty}\,,
	\end{split}\label{SCI:2}
\end{equation}
where we have introduced new parameters $y_\Psi=e^{\ri\pi\Delta_\Psi}$, $y_{T_r}=e^{\ri\pi\Delta_{T_r}}~(\Delta_\Psi,\Delta_{T_r}\in\mathbb R)$ and $\mn_\Psi,\mft_r\in\mathbb R$ via the relations\footnote{In \cite{Bobev:2022wem}, the ABJM SCI was analyzed in a similar fashion but the $(\Delta,\mn)$ parameters were introduced \emph{after} the change of integration variables involving the degrees of freedom originally carried by the fugacity associated with the topological symmetry. As a result, the ABJM SCI is written as a function of $(\Delta_\Psi,\mn_\Psi)$ only, which can take any configuration compatible with the superpotential charge constraints. We follow the same approach in Section~\ref{sec:ex} for other SCFT examples with 2 nodes, namely $N^{0,1,0}$ and $Q^{1,1,1}$ theories, so that $(\Delta_{T_r},\mft_r)$ do not carry any extra degrees of freedom.}
\begin{equation}
	y_\Psi q^{\mn_\Psi}=q^{R(\Psi)}\prod_x\xi_x^{f_x(\Psi)}\,,\qquad y_{T_r}q^{-\mft_r}=\xi_{T_r}\,.\label{SCI:redefine}
\end{equation}
We have also replaced the argument of the SCI accordingly,
\begin{equation}
	\mI(q,\boldsymbol{\xi})\quad\to\quad\mI(\omega,\Delta,\mn)\,,
\end{equation}
where $\Delta$ and $\mn$ collectively represent $\Delta=(\Delta_\Psi,\Delta_{T_r})$ and $\mn=(\mn_\Psi,\mft_r)$.

Next, using the property of the $\infty$-Pochhammer symbol (\ref{poch:inverse}) and assuming $q=e^{\ri\pi\omega}\in\mathbb{R}$, we can remove the absolute signs in (\ref{SCI:2}) and write
\begin{align}\label{SCI:3}
		&\mI(\omega,\Delta,\mn) \nonumber\\
		&=\fft{1}{(N!)^p}\sum_{\mm_1,\cdots,\mm_p\in\mathbb Z^N}\oint_{|s_{r,i}|=e^{-\ri\pi\omega\mm_{r,i}}}\bigg(\prod_{r=1}^p\prod_{i=1}^N\fft{ds_{r,i}}{2\pi \ri s_{r,i}}e^{\fft{\ri k_r}{4\pi\omega}(\bU_{r,i}^2-U_{r,i}^2)+\fft{\ri}{2\omega}(\Delta_{T_r}-\omega\mft_r)(\bU_{r,i}-U_{r,i})}\bigg)\nonumber\\
		&\quad\times(-1)^{\fft{pN(N-1)}{2}}\prod_{r=1}^p\prod_{i\neq j}^N(1-s_{r,i}^{-1}s_{r,j})^\fft12(1-\bs_{r,i}^{-1}\bs_{r,j})^\fft12\times \prod_{r=1}^p\prod_{i=1}^Ne^{-\fft{\ri}{\omega}\ell_{r,i}(\bU_{r,i}-U_{r,i})}\\
		&\quad\times\prod_{\Psi}\prod_{\rho_\Psi}e^{\fft{\ri}{8\pi\omega}(\rho_\Psi(\bU)^2-\rho_\Psi(U)^2)+\fft{\ri}{4\omega}\Delta_\Psi(\rho_\Psi(\bU)-\rho_\Psi(U))-\fft{\ri}{4}(1-\mn_\Psi)(\rho_\Psi(\bU)-\rho_\Psi(U))}\fft{(e^{-\ri\rho_\Psi(\bU)}y_{\Psi}^{-1}q^{2-\mn_\Psi};q^2)_\infty}{(e^{\ri\rho_\Psi(U)} y_\Psi q^{\mn_\Psi};q^2)_\infty}\,,\nonumber
\end{align}
where we have introduced new integration variables 
\begin{equation}
	\begin{split}
		s_{r,i}&=e^{\ri U_{r,i}}=z_{r,i}q^{-\mm_{r,i}}=e^{\ri(h_{r,i}-\pi\omega\mm_{r,i})}\,,\\
		\bs_{r,i}&=e^{-\ri\bU_{r,i}}=z_{r,i}^{-1}q^{-\mm_{r,i}}=e^{-\ri(h_{r,i}+\pi\omega\mm_{r,i})}\,,
	\end{split}\label{z:to:s}
\end{equation}
and also, for later convenience, introduced a trivial phase
\begin{equation}
	e^{-2\pi\ri \mm_{r,i}\ell_{r,i}}=e^{-\fft{\ri}{\omega}\ell_{r,i}(\bU_{r,i}-U_{r,i})}=1\qquad(\mm_{r,i},\ell_{r,i}\in\mathbb{Z})\,,\label{integer:phase}
\end{equation}
in terms of an arbitrary set of integers $\{\ell_{r,i}\}$. Importantly, the resulting expression of the SCI (\ref{SCI:3}) is factorized into holomorphic and anti-holomorphic parts with respect to the new integration variables (\ref{z:to:s}), as discussed in \cite{Choi:2019dfu, Choi:2019zpz} (see also \cite{Dimofte:2011py,Beem:2012mb,Hwang:2012jh}). \\

Before considering the Cardy-like limit of the factorized SCI (\ref{SCI:3}), we highlight a couple of technical steps entering the factorization of the $\mN=2$ SCI that differ from the previous analysis for the ABJM/ADHM SCI in our previous work \cite{Bobev:2022wem}:
\begin{itemize}
	\item The new integration variables (\ref{z:to:s}) are slightly different from the ones in \cite{Bobev:2022wem}. 
	
	\item The introduction of an arbitrary set of integers $\{\ell_{r,i}\}$ in (\ref{integer:phase}) is a new ingredient that streamlines the analysis. 
\end{itemize}
These two differences do not affect the value of the SCI, but they do change its factorized structure and thereby make the comparison between the Cardy-like expansion of the SCI and the TTI more straightforward, as we show below in Section~\ref{sec:general:relation}.

%%%%%
\subsubsection{Cardy-like limit}\label{sec:general:SCI:Cardy}
%%%%%
Now we take the Cardy-like limit
\begin{equation}
	q\to1^-\qquad\Leftrightarrow\qquad\omega\to \ri0^+\,.\label{Cardy-like}
\end{equation}
Following the details spelled out in Appendix \ref{app:Cardy}, one can expand the SCI (\ref{SCI:3}) in the Cardy-like limit (\ref{Cardy-like}) as \cite{Choi:2019dfu, Choi:2019zpz}
\begin{equation}
	\begin{split}
		\mI(\omega,\Delta,\mn)&=\fft{1}{(N!)^p}(-1)^{\fft{pN(N-1)}{2}}\int_{\mathbb C^{pN}}\bigg(\prod_{r=1}^p\prod_{i=1}^N\fft{dU_{r,i}d\bU_{r,i}}{-4\ri\pi^2\omega}\bigg)\\
		&\quad\times\exp[\fft{1}{\pi\omega}\Im\mW^{(0)}[U;\Delta,\ell]+2\Re\mW^{(1)}[U;\Delta,\mn]+\mO(\omega)]\,,
	\end{split}\label{SCI:Cardy:3}
\end{equation}
where we have introduced the Cardy-like expansion of a holomorphic effective potential,
\begin{subequations}
	\begin{align}
		\mW[U;\Delta,\mn,\omega,\ell]&=\mW^{(0)}[U;\Delta,\ell]+2\pi\ri\omega\,\mW^{(1)}[U;\Delta,\mn]+\mO(\omega^2)\,,\\
		\mW^{(0)}[U;\Delta,\ell]&=\sum_{r=1}^p\sum_{i=1}^N\bigg[\fft12k_rU_{r,i}^2-\pi(2\ell_{r,i}-\Delta_{T_r})U_{r,i}\bigg]\nn\\
		&\quad+\sum_\Psi\sum_{\rho_\Psi}\bigg[\fft14\rho_\Psi(U)^2+\fft{\pi\Delta_\Psi}{2}\rho_\Psi(U)-\text{Li}_2(e^{\ri\rho_\Psi(U)}y_\Psi)\bigg]\,,\label{mW:Cardy:0}\\
		\mW^{(1)}[U;\Delta,\mn]&=\fft{\ri}{2}\sum_{r=1}^p\mft_r\sum_{i=1}^NU_{r,i}-\fft12\sum_{r=1}^p\sum_{i\neq j}^N\text{Li}_1(e^{\ri(U_{r,j}-U_{r,i})})\nn\\
		&\quad+\fft{\ri}{4}\sum_\Psi(1-\mn_\Psi)\sum_{\rho_\Psi}\big(\rho_\Psi(U)+\pi\Delta_\Psi\big)\nn\\
		&\quad+\fft12\sum_{\Psi}(1-\mn_\Psi)\sum_{\rho_\Psi}\text{Li}_1(e^{\ri\rho_\Psi(U)}y_\Psi)\,.\label{mW:Cardy:1}
	\end{align}\label{mW:Cardy}%
\end{subequations}
Observe that the $\mathcal{O}(\omega^{-1})$ leading order effective potential (\ref{mW:Cardy:0}) depends on the set of arbitrary integers $\{\ell_{r,i}\}$ introduced previously. This fact plays a crucial role in matching the leading order effective potential for the SCI (\ref{mW:Cardy:0}) and the Bethe potential for the TTI introduced below in Section \ref{sec:general:relation}. 

Evaluating the integral (\ref{SCI:Cardy:3}) using the saddle-point approximation described in Appendix B of \cite{Bobev:2022wem}, we obtain the Cardy-like expansion of the log of the SCI
\begin{equation}
	\begin{split}
		\log\mI(\omega,\Delta,\mn)&=\fft{1}{\pi\omega}\Im\mW^{(0)}[U^\star;\Delta,\ell]+2\Re\mW^{(1)}[U^\star;\Delta,\mn]\\[1mm]
		&\quad-\fft12\log\det\mathbb{H}[U^\star;\Delta]+\log(-1)^{\fft{pN(N-1)}{2}}+\mO(\omega)\,,
	\end{split}\label{SCI:Cardy:saddle:1}
\end{equation}
where we have implicitly assumed that the contribution from a particular saddle point $\{U_{r,i}^\star\}$ satisfying the saddle point equation,
\begin{equation}
	\fft{\partial\mW^{(0)}[U;\Delta,\ell]}{\partial U_{r,i}}=0\,,\label{saddle:eq}
\end{equation}
yields a dominant contribution to the SCI in the Cardy-like limit. In (\ref{SCI:Cardy:saddle:1}), $\mathbb{H}$ denotes the Hessian matrix around the saddle point. Introducing $Y_I\in\{U_{1,i},\cdots,U_{p,N}\}$, its matrix components are given by
\begin{equation}
	\Big(\mathbb{H}[U;\Delta]\Big)_{I,J}=\begin{pmatrix}
			\mathbb{J}[U;\Delta] & 0 \\
			0 & -\overline{\mathbb{J}[U;\Delta]}
		\end{pmatrix}\,,\qquad
		\Big(\mathbb J[U;\Delta]\Big)_{I,J}\equiv\fft{\partial^2\mW^{(0)}[U;\Delta,\ell]}{\partial Y_I\partial Y_J}\,.\label{Hessian}
\end{equation}
Note that the $\ell$-dependence disappears in the 2nd derivative of the leading order effective potential $\mathcal{W}^{(0)}$. One can write down the Cardy-like expansion of the SCI (\ref{SCI:Cardy:saddle:1}) more explicitly as
\begin{equation}
	\begin{split}
		&\log\mI(\omega,\Delta,\mn)\\
		&=\fft{1}{\pi\omega}\Im\mW^{(0)}[U^\star;\Delta,\ell]\\
		&\quad+\log\Bigg|\fft{1}{\det\mathbb{J}}\prod_{r=1}^p\Bigg[\prod_{i=1}^Ns_{r,i}^{\mft_r}\prod_{i\neq j}^N\bigg(1-\fft{s_{r,i}}{s_{r,j}}\bigg)\Bigg]\prod_\Psi\prod_{\rho_\Psi}\bigg(\fft{e^{\ri\rho_\Psi(U)/2}y_\Psi^{1/2}}{1-e^{\ri\rho_\Psi(U)}y_\Psi}\bigg)^{1-\mn_\Psi}\Bigg|_{U=U^\star}\\
		&\quad+\underbrace{\log(-1)^{\fft{pN(N-1)}{2}}-\fft12\log(-1)^{pN}}_{\equiv\ri\varphi}+\mO(\omega)\,,
	\end{split}\label{SCI:Cardy:saddle:2}
\end{equation}
using the expression (\ref{mW:Cardy:1}) and the decomposition of the Hessian matrix (\ref{Hessian}). We collect the purely imaginary terms coming from the phase factors independent of flavor fugacities and magnetic fluxes and combine them into the quantity $\ri\varphi$ for notational convenience. This purely imaginary term depends only on the number of nodes $p$ and the rank of the gauge group at each node $N$, and in particular vanishes for even $p$.

%%%%%
\subsection{Topologically twisted index}\label{sec:general:TTI}
%%%%%
The TTI is defined as the partition function of a 3d $\mN=2$ theory on $S^1\times\Sigma_{\mg}$ with a partial topological twist on the Riemann surface $\Sigma_{\mg}$ that preserves two real supercharges.\footnote{In this work we take $\Sigma_{\mg}$ to be a compact Riemann surface of genus $\mathfrak{g}$ without punctures.} The TTI of $\mN=2$ supersymmetric gauge theories was first studied in \cite{Nekrasov:2014xaa} based on the Bethe/gauge correspondence \cite{Nekrasov:2009rc,Nekrasov:2009ui,Nekrasov:2009uh} and here we employ its matrix model representation obtained via supersymmetric localization \cite{Benini:2015noa,Benini:2016hjo,Closset:2016arn,Closset:2017zgf}. The main focus in this paper will be on the $\mathfrak{g}=0$ case, i.e. $\Sigma_{\mg} = S^2$. The matrix model for the class of $\mN=2$ SCFTs described in Section~\ref{sec:general:summary} reads \cite{Hosseini:2016tor,Hosseini:2016ume,PandoZayas:2020iqr,Bobev:2023lkx}\footnote{We absorb the extra phase factors in the TTI localization formula due to the periodic boundary condition for fermions along $S^1$ \cite{Closset:2017zgf} by redefining fugacities associated with the U(1) topological symmetries appropriately, see \cite{Benini:2015noa,Bobev:2022eus}.}
\begin{equation}
	\begin{split}
		Z(\Delta,\mn)&=\fft{1}{(N!)^p}\sum_{\mm_1,\ldots,\mm_p\in\mathbb Z^N}\int_{\mathcal C}\prod_{r=1}^p\Bigg[\prod_{i=1}^N\fft{dx_{r,i}}{2\pi \mathrm{i}x_{r,i}}(x_{r,i})^{k_r\mm_{r,i}+\mft_r}(y_{T_r})^{\mm_{r,i}}\prod_{i\neq j}^N\bigg(1-\fft{x_{r,i}}{x_{r,j}}\bigg)\Bigg] \\
		&\kern10em~\times\prod_\Psi\prod_{\rho_\Psi}\bigg(\fft{e^{\ri\rho_\Psi(u)/2}y_\Psi^{1/2}}{1-e^{\ri\rho_\Psi(u)}y_\Psi}\bigg)^{\rho_\Psi(\mm)-\mn_\Psi+1}\,, \label{TTI:1}
	\end{split}
\end{equation}
where the contour integrals for gauge zero modes $x_{r,i}=e^{\ri u_{r,i}}$ capture the Jeffrey-Kirwan (JK) residues, see \cite{Benini:2015eyy,Benini:2015noa,Closset:2016arn,Closset:2017zgf} for a detailed discussion of the contours. As in the matrix model for the SCI (\ref{SCI:1}), $\mm_{r,i}$ stand for quantized gauge magnetic fluxes and $\rho_\Psi$ runs over the weights of the representation of the $\mathcal N=2$ chiral multiplet $\Psi$ under the gauge group. The fugacities $y_\Psi=e^{\ri\pi\Delta_\Psi}$, $y_{T_r}=e^{\ri\pi\Delta_{m,r}}$ and the background magnetic fluxes $\mn_\Psi$, $\mft_r$ are associated with flavor symmetries of a given theory. We omit $N$ and $\{k_r\}$ in the argument of the TTI for notational convenience throughout this paper. \\

The matrix model for the TTI (\ref{TTI:1}) can be evaluated directly using the Bethe Ansatz (BA) formalism \cite{Benini:2015eyy,Hosseini:2016tor}, which we now briefly review. The first step is to implement the sum over gauge magnetic fluxes in (\ref{TTI:1}) as
\begin{equation}
	\begin{split}
		Z(\Delta,\mn)&=\fft{1}{(N!)^p}\int_{\mathcal C}\prod_{r=1}^p\Bigg[\prod_{i=1}^N\fft{dx_{r,i}}{2\pi \mathrm{i}x_{r,i}}x_{r,i}^{\mft_r}\fft{e^{\ri MB_{r,i}}}{e^{\ri B_{r,i}}-1}\prod_{i\neq j}^N\bigg(1-\fft{x_{r,i}}{x_{r,j}}\bigg)\Bigg] \\
		&\quad\times\prod_\Psi\prod_{\rho_\Psi}\bigg(\fft{e^{\ri\rho_\Psi(u)/2}y_\Psi^{1/2}}{1-e^{\ri\rho_\Psi(u)}y_\Psi}\bigg)^{1-\mn_\Psi}\,, \label{TTI:2}
	\end{split}
\end{equation}
where we have introduced a large positive integer $M$\footnote{This cutoff $M$ is introduced to capture the JK residues appropriately, see \cite{Benini:2015eyy,Hosseini:2016tor,Hosseini:2016ume,PandoZayas:2020iqr,Bobev:2023lkx} for example. For a vanishing CS level one can take $k_r\to0^\pm$ for the choice of $\text{sign}[k_r]$, which has been done implicitly for various examples in \cite{Hosseini:2016ume}. Note that the BAE (\ref{BAE:simple}) and the corresponding BA formula (\ref{TTI:3}) are ultimately insensitive to the direction of the limit. In Section \ref{sec:ex} we take $k_1\to0^+$ for the $V^{5,2}$ theory and  $k_1=-k_2=k\to0^+$ for the $Q^{1,1,1}$ theory.} and the BA operators $B_{r,i}$ via
\begin{equation}
		e^{\ri\,\text{sign}[k_r]B_{r,i}}=y_{T_r}x_{r,i}^{k_r}\prod_\Psi\prod_{\rho_\Psi}\bigg(\fft{e^{\ri\rho_\Psi(u)/2}y_\Psi^{1/2}}{1-e^{\ri\rho_\Psi(u)}y_\Psi}\bigg)^{(\rho_\Psi)_{r,i}}\,.\label{BA:operator}
\end{equation}
We refer to Appendix \ref{app:SCI-phase} for the definition of the symbol $(\rho_\Psi)_{r,i}$ for a given weight $\rho_\Psi$. 

Next, replacing the integrals in (\ref{TTI:2}) with the sum over solutions to the Bethe Ansatz Equations (BAE),
\begin{equation}
	e^{\ri\,\text{sign}[k_r]B_{r,i}}=1\,,\label{BAE:simple}
\end{equation}
we obtain the BA formula 
\begin{equation}
	\begin{split}
		Z(\Delta,\mn)=\sum_{\{u_{r,i}\}\in\text{BAE}}\fft{1}{\det\mathbb{B}}\prod_{r=1}^p\Bigg[\prod_{i=1}^Nx_{r,i}^{\mft_r}\prod_{i\neq j}^N\bigg(1-\fft{x_{r,i}}{x_{r,j}}\bigg)\Bigg]\prod_\Psi\prod_{\rho_\Psi}\bigg(\fft{e^{\ri\rho_\Psi(u)/2}y_\Psi^{1/2}}{1-e^{\ri\rho_\Psi(u)}y_\Psi}\bigg)^{1-\mn_\Psi}\,,\label{TTI:3}
	\end{split}
\end{equation}
where the Jacobian matrix is defined as
\begin{equation}
	\mathbb{B}\equiv\fft{\partial(e^{\ri B_{1,1}},\cdots,e^{\ri B_{1,N}},\cdots,e^{\ri B_{p,1}},\cdots e^{\ri B_{p,N}})}{\partial(\log x_{1,1},\cdots,\log x_{1,N},\cdots,\log x_{p,1},\cdots,\log x_{p,N})}\,.\label{def:B}
\end{equation}
Note that the factor of $(N!)^{-p}$ in (\ref{TTI:2}) is canceled by the degeneracy of a particular BAE solution from permutations in the BA formula (\ref{TTI:3}). Taking the log, the BA formula (\ref{TTI:3}) can be written as
\begin{align}\label{TTI:4}
		\log Z(\Delta,\mn)&=\log\Bigg[\fft{1}{\det\mathbb{B}}\prod_{r=1}^p\Bigg[\prod_{i=1}^Nx_{r,i}^{\mft_r}\prod_{i\neq j}^N\bigg(1-\fft{x_{r,i}}{x_{r,j}}\bigg)\Bigg]\prod_\Psi\prod_{\rho_\Psi}\bigg(\fft{e^{\ri\rho_\Psi(u)/2}y_\Psi^{1/2}}{1-e^{\ri\rho_\Psi(u)}y_\Psi}\bigg)^{1-\mn_\Psi}\Bigg]_{u=u^\star}\nonumber \\[1mm]
		&\quad+(\text{contribution from other BAE solutions})\,.
\end{align}
Note that in this expression we have emphasized the contribution to the TTI from a particular BAE solution $\{u_{r,i}^\star\}$ of the equations in \eqref{BAE:simple}.\\

An important fact about the BA formulation of the TTI is that the BAE can be derived from extremizing a single function, as follows. We first rewrite the BAE (\ref{BAE:simple}) more explicitly by taking the logarithm of (\ref{BA:operator}) as
\begin{equation}
	2\pi\,\text{sign}[k_r]n_{r,i}=\pi\Delta_{m,r}+k_ru_{r,i}-\ri\sum_\Psi\sum_{\rho_\Psi}(\rho_\Psi)_{r,i}\bigg[\text{Li}_1(e^{\ri\rho_\Psi(u)}y_\Psi)+\fft{\ri\rho_\Psi(u)}{2}+\fft{\ri\pi\Delta_\Psi}{2}\bigg]\,,\label{BAE}
\end{equation}
for an arbitrary set of integers $n_{r,i}\in\mathbb Z$ that comes from the ambiguity $e^{2\ri\pi\mathbb{Z}}=1$. The Bethe potential is then introduced as
\begin{equation}
\begin{split}
	\mV[u;\Delta,\mn]&=\sum_{r=1}^p\sum_{i=1}^N\bigg[-\fft{k_r}{2}u_{r,i}^2+\pi(2\,\text{sign}[k_r]n_{r,i}-\Delta_{m,r})u_{r,i}\bigg]\\
	&\quad+\sum_{\Psi}\sum_{\rho_\Psi}\bigg[\text{Li}_2(e^{\ri\rho_\Psi(u)}y_\Psi)-\fft14\rho_\Psi(u)^2-\fft{\pi\Delta_\Psi}{2}\rho_\Psi(u)\bigg]\,,
\end{split}\label{V}
\end{equation}
which yields the BAE (\ref{BAE}) upon extremizing with respect to the gauge holonomy $u_{r,i}$. Note that in (\ref{V}) we have implicitly fixed the integration constant. The Bethe potential (\ref{V}) can also be understood as the effective twisted superpotential that governs the low-energy dynamics on the Coulomb branch of the 2d $A$-twist $\mN=(2,2)$ theory on the Riemann surface \cite{Closset:2016arn,Closset:2017zgf} that we have chosen here to be $\Sigma_{\mathfrak{g}}=S^2$.

%%%%%
\subsection{Relation between indices}\label{sec:general:relation}
%%%%%
In this subsection, we derive the Bethe formulation of the SCI (\ref{summary}) in a more precise form based on the Cardy-like expansion of the SCI in Section~\ref{sec:general:SCI} and the BA formulation of the TTI in Section~\ref{sec:general:TTI}. \\

The key observation is that the leading order effective potential for the SCI in the Cardy-like limit (\ref{mW:Cardy:0}) is precisely the same as (minus) the Bethe potential for the TTI (\ref{V}), 
\begin{equation}
	\mW^{(0)}[U;\Delta,\ell] \quad\leftrightarrow\quad -\mV[u;\Delta,\mn]\,,\label{mW=-V}
\end{equation}
provided we identify various parameters as
\begin{equation}
	\begin{split}
		\text{SCI parameters}\kern2em\,&\kern4em\quad~\text{TTI parameters}\\
		(U_{r,i}\,,\, \Delta_{T_r}\,,\, \Delta_\Psi\,,\, \ell_{r,i})\quad&\leftrightarrow\quad (u_{r,i}\,,\, \Delta_{m,r}\,,\, \Delta_\Psi\,,\, \text{sign}[k_r]n_{r,i})\,.
	\end{split}\label{map:1}
\end{equation}
As a consequence of this identification, the saddle point equation for the SCI (\ref{saddle:eq}) and the BAE for the TTI (\ref{BAE}) become equivalent,
\begin{equation}
	\fft{\partial \mW^{(0)}[U;\Delta,\ell]}{\partial U_{r,i}}=0\qquad\leftrightarrow\qquad\fft{\partial \mV[u;\Delta,\mn]}{\partial u_{r,i}}=0\,,
\end{equation}
and therefore the saddle point solution becomes equivalent to the BAE solution:
\begin{equation}
	\{U_{r,i}^\star\}\quad\leftrightarrow\quad\{u_{r,i}^\star\}\,.\label{map:2}
\end{equation}
The Hessian $\mathbb{H}$ for the SCI saddle point approximation (\ref{Hessian}) and the Jacobian $\mathbb{B}$ for the TTI BA formula (\ref{def:B}) are also related under the identifications (\ref{mW=-V}) and (\ref{map:1}) as
\begin{equation}
		s_I\big(\mathbb{J}\big)_{I,J}\bigg|_{U=U^\star} = s_I\fft{\partial^2\mW^{(0)}[U;\Delta,\ell]}{\partial Y_I\partial Y_J}\bigg|_{U=U^\star}~~\leftrightarrow~~-\big(\mathbb{B}\big)_{I,J}\bigg|_{u=u^\star}=-s_I\fft{\partial^2\mV[u;\Delta,\mn]}{\partial y_I\partial y_J}\bigg|_{u=u^\star}\,,\label{B:J}
\end{equation}
in terms of $y_I=u_{r,i}$ and some phases $s_I=-\text{sign}[k_r]$, which can be derived from
\begin{equation}
	B_{r,i}=-\text{sign}[k_r]\fft{\partial\mV[u;\Delta,\mn]}{\partial u_{r,i}}\,.
\end{equation}
From (\ref{B:J}), we can identify the absolute values of the determinants of those matrices as
\begin{equation}
	\big\vert\det\mathbb{B}\;\big\vert_{u=u^\star}\quad\leftrightarrow\quad\big\vert\det\mathbb{J}\;\big\vert_{U=U^\star}\,.\label{map:3}
\end{equation}
Finally, the parameters $(\mn_\Psi,\mft_r)$ used in both indices are identified in a straightforward way. \\

Rewriting the Cardy-like expansion of the SCI (\ref{SCI:Cardy:saddle:2}) in terms of the Bethe potential (\ref{V}) and the BA formula for the TTI (\ref{TTI:4}) based on the above identifications, we find
\begin{empheq}[box=\fbox]{equation}
	\begin{split}
		\log\mI(\omega,\Delta,\mn)=-\fft{1}{\pi\omega}\Im\mV[u^\star;\Delta,\mn]+\Re\log Z(\Delta,\mn)+\ri\varphi+\mO(\omega)\,,
	\end{split}\label{summary:improved}
\end{empheq}
which completes the derivation of the relation (\ref{summary}) between the two indices. We thus conclude that the leading and first subleading orders in the Cardy-like expansion of the SCI are governed by the Bethe potential and the TTI of the same $\mN=2$ SCFT, respectively. Before exhibiting this relation in concrete examples of 3d $\mathcal{N}=2$ SCFTs, we collect a number of comments below.
\begin{itemize}
	\item The relation (\ref{summary:improved}) is derived under the Cardy-like limit of the SCI but does not involve any other limit such as the large $N$ limit for holographic SCFTs. Hence the identification of the first two leading terms of the SCI in the Cardy-like expansion with the Bethe potential and the TTI is valid non-perturbatively in the large $N$ expansion of holographic SCFTs and at finite $N$ \cite{Choi:2019zpz}. We will discuss this further in Section \ref{sec:ex}.
	
	\item The Bethe potential evaluated at the BAE solution $\{u^\star\}$ in the r.h.s. of (\ref{summary:improved}) does not depend on the set of arbitrary integers $\{n_{r,i}\}$ introduced in the BAE (\ref{BAE}), which is indeed expected since they are unphysical and simply come from the $2\pi\ri\mathbb{Z}$ ambiguity in the exponent of the BA operators (\ref{BAE:simple}).
	
	\item In the presentation above, we have assumed that a particular saddle point (resp. BAE solution) yields a dominant contribution to the SCI (resp. TTI) in deriving the index relation (\ref{summary:improved}). However, since the relation between the leading order effective potential and the Bethe potential given in (\ref{mW=-V}) is valid before specifying a saddle point or a BAE solution, one can easily generalize the index relation (\ref{summary:improved}) to the contribution from each saddle point and BAE solution. To be more specific, if we allow for multiple contributions from different saddle points $\{u^\star_{(\sigma)}\}$ and BAE solutions $\{U^\star_{(\sigma)}\}$ labeled by $\sigma$ as
	\begin{equation}
	\begin{split}
		\mI(\omega,\Delta,\mn)&=\sum_{\sigma}\mI_{(\sigma)}(\omega,\Delta,\mn)\,,\\
		Z(\Delta,\mn)&=\sum_{\sigma}Z_{(\sigma)}(\Delta,\mn)\,,
	\end{split}
	\end{equation}
	based on the expressions (\ref{SCI:Cardy:3}) and (\ref{TTI:3}) respectively, we obtain
	\begin{equation}
			\log\mI_{(\sigma)}(\omega,\Delta,\mn)=-\fft{1}{\pi\omega}\Im\mV[u^\star_{(\sigma)};\Delta,\mn]+\Re\log Z_{(\sigma)}(\Delta,\mn)+\ri\varphi+\mO(\omega)\,.\label{summary:split}
	\end{equation}
	This generalizes the index relation (\ref{summary:improved}) to multiple saddles.
	
	\item The parameters $(\mn_\Psi,\mft_r)$ represent the background magnetic fluxes for flavor symmetries in the TTI. The same parameters in the SCI do not have the same physical meaning: they are simply introduced by redefining flavor fugacities as (\ref{SCI:redefine}). In particular, we did not turn on background magnetic fluxes for flavor symmetries in the matrix model for the SCI (\ref{SCI:1}). The generalized SCI involving background magnetic fluxes can also be written in terms of a matrix model \cite{Kapustin:2011jm} and its Cardy-like limit was also studied in \cite{Choi:2019dfu}. The fact that the parameters $(\mn_\Psi,\mft_r)$ in the SCI are precisely identified with the background magnetic fluxes in the TTI through the Bethe formulation of the SCI (\ref{summary:improved}) therefore requires further explanation. We plan on studying the physics behind this parameter matching and investigating the index relation (\ref{summary:improved}) for a generalized SCI in future work.
	
	\item The identification between parameters (\ref{map:1}) is not the same as the one proposed for the ABJM/ADHM theories in \cite{Bobev:2022wem}. This is because we have improved a couple of technical steps in the factorization of the SCI (see Section~\ref{sec:general:SCI:fact}) and thereby the holomorphic effective potential in the Cardy-like limit (\ref{mW:Cardy}) is slightly different from the one used in \cite{Bobev:2022wem}. The final Cardy-like expansion of the SCI for the ABJM/ADHM theories obtained in the generic $\mN=2$ conventions of the present paper are ultimately equivalent to the results of \cite{Bobev:2022wem}, and we show this explicitly in Appendix~\ref{app:ABJMADHM}.
	
	\item The Bethe potential and the TTI of a single chiral multiplet $\Psi$ with $R$-charge $r$ and a U(1) flavor charge $f$ are indeed given by (\ref{VTTI:single:1}). To see this explicitly, one may read off the Bethe potential and the TTI for a single chiral multiplet from (\ref{V}) and (\ref{TTI:1}) as
	\begin{equation}
		\mV_\Psi=\text{Li}_{2}(y_\Psi)\,,\qquad\text{and}\qquad Z_\Psi=\bigg(\fft{y_\Psi^{1/2}}{1-y_\Psi}\bigg)^{1-\mn_\Psi}\,,\label{VTTI:single:2}
	\end{equation}
	respectively. This is precisely the same as (\ref{VTTI:single:1}) under the identification of SCI/TTI parameters (\ref{map:1}) and the reparametrization (\ref{SCI:redefine}), modulo minor notational differences; in (\ref{VTTI:single:1}), we have simply used a subscript ``$i$'' to present fugacities/charges associated with a chiral multiplet $\Psi_i$. Note that the reparametrization for a single chiral multiplet (\ref{SCI:redefine:single}) simply corresponds to a special case of the generic one (\ref{SCI:redefine}). 
	
	\item The relation between the SCI and the TTI (\ref{summary:improved}) is clearly different from the relation between the $A$-twist partition function and the TTI recently observed in \cite{Amariti:2023ygn}. The latter is motivated by describing the 3d $\mN=2$ theory on a Seifert manifold in terms of the $A$-twist of the 2d $\mN=(2,2)$ theory, associated with the parent 3d $\mathcal{N}=2$ theory after a circle reduction, on the Coulomb branch based on the work of \cite{Closset:2016arn,Closset:2017zgf,Closset:2018ghr,Closset:2019hyt}. The supersymmetric background for the SCI, however, does not admit any Seifert structure \cite{Closset:2018ghr,Closset:2019hyt} and therefore our index relation (\ref{summary:improved}) cannot be derived relying solely on the framework of \cite{Closset:2016arn,Closset:2017zgf,Closset:2018ghr,Closset:2019hyt}. In fact, our observation (\ref{summary:improved}) suggests that there are interesting relations between 3d supersymmetric partition functions that go beyond the cases where the backgrounds allow for a Seifert description.
\end{itemize}
%

%%%%%
\section{3d holographic SCFTs from M2-branes }\label{sec:ex}
%%%%%
We now employ the relation (\ref{summary:improved}) to study the SCI of various 3d $\mN=2$ holographic SCFTs arising on the worldvolume of $N$ coincident M2-branes probing orbifold singularities. To be specific, we present the all-order $1/N$ perturbative expansion for the first two leading terms of the SCI in the Cardy-like limit. This is achieved by applying recent numerical techniques to study the corresponding TTI \cite{Bobev:2022jte,Bobev:2022eus,Bobev:2023lkx} and using the resulting expressions in conjunction with the relation (\ref{summary:improved}). We consider 3d $\mN=2$ SCFTs holographically dual to M-theory on AdS$_4\times Y_7$ for three Sasaki-Einstein orbifolds $Y_7\in\{N^{0,1,0}/\mathbb{Z}_k,V^{5,2}/\mathbb{Z}_{N_f},Q^{1,1,1}/\mathbb{Z}_{N_f}\}$. This analysis mirrors a similar approach previously used for ABJM/ADHM theories \cite{Bobev:2022wem}. 

%%%%%
\subsection{$N^{0,1,0}$ theory}\label{sec:ex:N010}
%%%%%

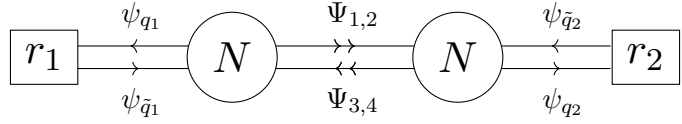
\begin{figure}
	\centering
	\begin{tikzpicture}
		\draw[->-=0.47,->-=0.57] (0.57,0.15) -- (2.43,0.15);
		\draw[->-=0.47,->-=0.57] (2.43,-0.15) -- (0.57,-0.15);
		\draw[->-=0.5] (-0.57,0.15) -- (-2.13,0.15);
		\draw[->-=0.57] (-2.13,-0.15) -- (-0.57,-0.15);
		\draw[->-=0.5] (5.03,0.15) -- (3.47,0.15);
		\draw[->-=0.57] (3.47,-0.15) -- (5.03,-0.15);
		\node at (-2.5,0) [rectangle,draw,scale=1.5,fill=white] {$r_1$};
		\node at (0,0) [circle,draw,scale=1.5,fill=white] {$N$};
		\node at (3,0) [circle,draw,scale=1.5,fill=white] {$N$};
		\node at (5.5,0) [rectangle,draw,scale=1.5,fill=white] {$r_2$};
		\node at (1.6,0.55) {$\Psi_{1,2}$};
		\node at (1.6,-0.6) {$\Psi_{3,4}$};
		\node at (-1.2,0.55) {$\psi_{q_1}$};
		\node at (-1.2,-0.6) {$\psi_{\tq_1}$};
		\node at (4.4,0.55) {$\psi_{\tq_2}$};
		\node at (4.4,-0.6) {$\psi_{q_2}$};
	\end{tikzpicture}
	\caption{Quiver diagram for the $N^{0,1,0}$ theory}\label{quiver:N010}
\end{figure}

We start with the 3d $\mN=2$ SCFT dual to M-theory on AdS$_4\times N^{0,1,0}/\mathbb{Z}_k$, which we simply refer to as the $N^{0,1,0}$ theory. We refer the reader to \cite{Gaiotto:2009tk,Hohenegger:2009as,Hikida:2009tp,Cheon:2011th,Hosseini:2016ume,Bobev:2023lkx} for more details about various aspects of this SCFT. The CS-matter theory in the UV is described by the quiver diagram shown in Fig.~\ref{quiver:N010}. In intermediate stages of the calculation we keep the number of fundamental and anti-fundamental pairs of chiral multiplets $r=r_1+r_2$ independent from the CS level $k$ as in \cite{Bobev:2023lkx}, although we stress that one needs to impose $r=k$ to correctly describe the $N^{0,1,0}$ theory \cite{Cheon:2011th,Bobev:2023lkx}.

To study the $N^{0,1,0}$ SCI using the index relation (\ref{summary:improved}), we leverage the results obtained for the $N^{0,1,0}$ TTI presented in \cite{Bobev:2023lkx}. To facilitate this approach, we first align the generic $\mN=2$ TTI conventions outlined in Section~\ref{sec:general:TTI} with those used in \cite{Bobev:2023lkx}. 

The Bethe potential (\ref{V}) can be written explicitly for the $N^{0,1,0}$ theory as
\begin{align}\label{V:N010}
		&\mV_{N^{0,1,0}}[u;\Delta,\mn] \nonumber\\
		&=\sum_{i=1}^N\bigg[-\fft{k}{2}(u_{1,i}^2-u_{2,i}^2)+\pi(2n_{1,i}-\Delta_{m,1})u_{1,i}+\pi(-2n_{2,i}-\Delta_{m,2})u_{2,i}\bigg] \nonumber\\
		&\quad+\sum_{a=1}^2\sum_{i,j=1}^N\bigg[\text{Li}_2(e^{\ri(u_{1,i}-u_{2,j}+\pi\Delta_a)})-\fft14(u_{1,i}-u_{2,j})^2-\fft{\pi\Delta_a}{2}(u_{1,i}-u_{2,j})\bigg] \nonumber\\
		&\quad+\sum_{a=3}^4\sum_{i,j=1}^N\bigg[\text{Li}_2(e^{\ri(u_{2,j}-u_{1,i}+\pi\Delta_a)})-\fft14(u_{2,j}-u_{1,i})^2-\fft{\pi\Delta_a}{2}(u_{2,j}-u_{1,i})\bigg]\\
		&\quad+r_1\sum_{i=1}^N\bigg[\text{Li}_2(e^{\ri(u_{1,i}+\pi\Delta_{q_1})})-\fft14u_{1,i}^2-\fft{\pi\Delta_{q_1}}{2}u_{1,i}+\text{Li}_2(e^{\ri(-u_{1,i}+\pi\Delta_{\tq_1})})-\fft14u_{1,i}^2+\fft{\pi\Delta_{\tq_1}}{2}u_{1,i}\bigg]\nonumber\\
		&\quad+r_2\sum_{i=1}^N\bigg[\text{Li}_2(e^{\ri(u_{2,i}+\pi\Delta_{q_2})})-\fft14u_{2,i}^2-\fft{\pi\Delta_{q_2}}{2}u_{2,i}+\text{Li}_2(e^{\ri(-u_{2,i}+\pi\Delta_{\tq_2})})-\fft14u_{2,i}^2+\fft{\pi\Delta_{\tq_2}}{2}u_{2,i}\bigg]\nonumber\,.
\end{align}
The BA formula for the TTI (\ref{TTI:3}) can be written explicitly for the $N^{0,1,0}$ theory and reads
\begin{align}\label{TTI:N010}
		Z_{N^{0,1,0}}(\Delta,\mn)&=\sum_{\{u_{r,i}\}\in\text{BAE}}\fft{1}{\det\mathbb{B}}\prod_{r=1}^2\Bigg[\prod_{i=1}^Nx_{r,i}^{\mft_r}\prod_{i\neq j}^N\bigg(1-\fft{x_{r,i}}{x_{r,j}}\bigg)\Bigg]\nonumber\\
		&\quad\times\prod_{i,j=1}^N\Bigg[\prod_{a=1}^2\bigg(\fft{e^{\ri(u_{1,i}-u_{2,j}+\pi\Delta_a)/2}}{1-e^{\ri(u_{1,i}-u_{2,j}+\pi\Delta_a)}}\bigg)^{1-\mn_a}\times\prod_{a=3}^4\bigg(\fft{e^{\ri(u_{2,j}-u_{1,i}+\pi\Delta_a)/2}}{1-e^{\ri(u_{2,j}-u_{1,i}+\pi\Delta_a)}}\bigg)^{1-\mn_a}\Bigg]\nonumber\\
		&\quad\times\prod_{i=1}^N\Bigg[\bigg(\fft{e^{\ri(u_{1,i}+\pi\Delta_{q_1})/2}}{1-e^{\ri(u_{1,i}+\pi\Delta_{q_1})}}\bigg)^{1-\mn_{q_1}}\bigg(\fft{e^{\ri(-u_{1,i}+\pi\Delta_{\tq_1})/2}}{1-e^{\ri(-u_{1,i}+\pi\Delta_{\tq_1})}}\bigg)^{1-\mn_{\tq_1}}\Bigg]^{r_1}\\
		&\quad\times\prod_{i=1}^N\Bigg[\bigg(\fft{e^{\ri(u_{2,i}+\pi\Delta_{q_2})/2}}{1-e^{\ri(u_{2,i}+\pi\Delta_{q_2})}}\bigg)^{1-\mn_{q_2}}\bigg(\fft{e^{\ri(-u_{2,i}+\pi\Delta_{\tq_2})/2}}{1-e^{\ri(-u_{2,i}+\pi\Delta_{\tq_2})}}\bigg)^{1-\mn_{\tq_2}}\Bigg]^{r_2}\nonumber\,,
\end{align}
where the flavor chemical potentials and magnetic fluxes are constrained as
\begin{equation}
	\begin{alignedat}{2}
		1&=\Delta_1+\Delta_4=\Delta_2+\Delta_3\,,&\qquad1&=\Delta_{q_1}+\Delta_{\tq_1}=\Delta_{q_2}+\Delta_{\tq_2}\,,\\
		1&=\mn_1+\mn_4=\mn_2+\mn_3\,,&\qquad 1&=\mn_{q_1}+\mn_{\tq_1}=\mn_{q_2}+\mn_{\tq_2}\,.
	\end{alignedat}\label{N010:constraints}
\end{equation}
As shown in Appendix \ref{app:Bethe:N010}, the expressions (\ref{V:N010}) and (\ref{TTI:N010}) given in the generic $\mN=2$ TTI conventions indeed match the Bethe potential of the $N^{0,1,0}$ theory presented in \cite{Bobev:2023lkx}. Given this, we do not need to solve the BAE obtained by differentiating the Bethe potential (\ref{V:N010}) from scratch and can directly employ the numerical BAE solutions constructed in \cite{Bobev:2023lkx} instead.\footnote{These solve the $N^{0,1,0}$ BAE derived from \eqref{V:N010} for a choice of integers $(n_{1,i},n_{2,i})=(1-i+N,i)$.} Note that the results in \cite{Bobev:2023lkx} were obtained for a symmetric quiver with $r_1=r_2=r/2$ and focusing on the so-called superconformal configuration 
\begin{equation}
	\Delta_a=\Delta_{q_{1,2}}=\Delta_{\tq_{1,2}}=\fft12\,,\qquad \mn_a=\mn_{q_{1,2}}=\mn_{\tq_{1,2}}=\fft12\,,\label{N010:special}
\end{equation}
Therefore, our study of the index relation will be limited to these configurations which we collectively denote by $(\Delta_{\text{sc}}, \mn_{\text{sc}})$ in what follows. The exact $N^{0,1,0}$ TTI deduced by substituting these numerical BAE solutions into the BA formula (\ref{TTI:N010}) then reads \cite{Bobev:2023lkx}
\begin{equation}
	\begin{split}
		\Re\log Z_{N^{0,1,0}}(\Delta_{\text{sc}},\mn_{\text{sc}}) &=-\fft{2\pi(k+r)}{3\sqrt{2k+r}}\left((\hat N_{k,r})^\fft32-\left(\fft{r}{4}+\fft{3k+2r}{(k+r)^2}\right)(\hat N_{k,r})^\fft12\right)\\[1mm]
		&\quad-\fft12\log\hat N_{k,r}+\hat f_0(k,r)+\hat f_\text{np}(N,k,r)\, ,
	\end{split}\label{TTI:N010:result}
\end{equation}
where the shifted $N$ parameter is given by
\begin{equation}
	\hat N_{k,r}=N+\fft{7r-2k}{48}+\fft{2}{3(k+r)}\,.\label{N010:details}
\end{equation}
The non-perturbative correction in (\ref{TTI:N010:result}) is exponentially suppressed at large $N$, i.e. $\hat f_\text{np}(N,k,r)\sim\mO(e^{-\sqrt{N}})$. Numerical values of the $N$-independent contribution $\hat f_0(k,r)$ are presented in \cite{Bobev:2023lkx} for various configurations of $(k,r)$ and below we provide some of those values. 
\begin{center}
	\footnotesize
	\renewcommand*{\arraystretch}{1.2}
	\begin{tabular}{ |c|c||c|c| }
		\hline
		$(k,r)$ & $\hat f_0(k,r)$ & $(k,r)$ & $\hat f_0(k,r)$ \\
		\hline\hline
		$(1,1)$ & $-2.2479735914758641588$ & $(2,3)$ & $-2.1883848791741933989$ \\
		\hline
		$(1,2)$ & $-2.2917317046811495268$ & $(3,2)$ & $-1.6284176444001315906$ \\
		\hline
		$(4,2)$ & $-1.6979156145862367914$ & $(2,6)$ & $-4.6689712958005925271$ \\ 
		\hline
	\end{tabular}
\end{center}

Employing the index relation (\ref{summary:improved}), the $N^{0,1,0}$ TTI result (\ref{TTI:N010:result}) determines the $\omega^0$ order coefficient of the SCI in the Cardy-like limit. To determine the $\omega^{-1}$ leading order coefficient of the SCI, we numerically evaluate the Bethe potential (\ref{V:N010}) for various BAE solutions $\{u_\star\}$ found in \cite{Bobev:2023lkx} and based on the very precise numerical data propose a closed form analytic expression. The result is given by
\begin{equation}
	\Im\mV_{N^{0,1,0}}[u_\star;\Delta_{\text{sc}},\mn] = 2\pi\Bigg[\fft{\pi(k+r)}{6\sqrt{2k+r}}\hat N_{k,r}^\fft32+\hat g_0(k,r)+\hat g_\text{np}(N,k,r)\Bigg]\,,\label{mW:Cardy:0:N010:saddle} 
\end{equation}
for the $\Delta$-configuration (\ref{N010:special}). Similarly to the TTI above, the correction $\hat g_{\text{np}}$ is exponentially suppressed in the large $N$ limit and we have only been able to determine the $N$-independent term $\hat g_0$ numerically. Below we present numerical values of $\hat g_0$ in selected examples.
\begin{center}
	\footnotesize
	\renewcommand*{\arraystretch}{1.2}
	\begin{tabular}{ |c|c||c|c| }
		\hline
		$(k,r)$ & $\hat g_0(k,r)$ & $(k,r)$ & $\hat g_0(k,r)$ \\
		\hline\hline
		$(1,1)$ & $-0.1601419698035751424$ & $(2,3)$ & $-0.3315699873474639015$ \\
		\hline
		$(1,2)$ & $-0.2224298877692041968$ & $(3,2)$ & $-0.2159504219974937479$ \\
		\hline
		$(4,2)$ & $-0.2261565608197277915$ & $(2,6)$ & $-1.0105562008492962901$ \\ 
		\hline
	\end{tabular}
\end{center}
We refer to Appendix \ref{app:Bethe:N010} for the numerical data that supports the analytic expression (\ref{mW:Cardy:0:N010:saddle}). \\

Substituting the TTI (\ref{TTI:N010:result}) and the Bethe potential (\ref{mW:Cardy:0:N010:saddle}) into the Cardy-like expansion~(\ref{summary:improved}), we obtain the SCI for the $N^{0,1,0}$ theory in the Cardy-like limit 
\begin{equation}
	\begin{split}
		\log\mI_{N^{0,1,0}}(\omega,\Delta_{\text{sc}},\mn_{\text{sc}})&=-\fft{2}{\omega}\Bigg[\fft{\pi(k+r)}{6\sqrt{2k+r}}\hat N_{k,r}^\fft32+\hat g_0(k,r)+\hat g_\text{np}(N,k,r)\Bigg]\\
		&\quad+\Bigg[-\fft{2\pi(k+r)}{3\sqrt{2k+r}}\left((\hat N_{k,r})^\fft32-\left(\fft{r}{4}+\fft{3k+2r}{(k+r)^2}\right)(\hat N_{k,r})^\fft12\right)\\
		&\kern3em-\fft12\log\hat N_{k,r}+\hat f_0(k,r)+\hat f_\text{np}(N,k,r)\Bigg]+\mathcal O(\omega)\,.
	\end{split}\label{SCI:N010}
\end{equation}
Note that the large $N$ non-perturbative corrections of order $\mO(e^{-\sqrt{N}})$ in the first two leading terms above are determined unambiguously by the corresponding corrections to the Bethe potential and the TTI, respectively.

%%%%%
\subsection{$V^{5,2}$ theory}\label{sec:ex:V52}
\begin{figure}
	\centering
	\begin{tikzpicture}
		\draw[->-=0.52] (0.57,0.15) -- (2.43,0.15);
		\draw[->-=0.52] (2.43,-0.15) -- (0.57,-0.15);
		\draw[->-=0.45,->-=0.5,->-=0.55] (-0.2,0) arc (0:360:0.8) ;
		\node at (0,0) [circle,draw,scale=1.5,fill=white] {$N$};
		\node at (2.85,0) [rectangle,draw,scale=1.2,fill=white] {$N_f$};
		\node at (-2.3,0) {$\Psi_I$};
		\node at (1.55,0.5) {$\psi_q$};
		\node at (1.55,-0.6) {$\psi_{\tq}$};
	\end{tikzpicture}
	\caption{Quiver diagram for the $V^{5,2}$ theory}\label{quiver:V52}
\end{figure}
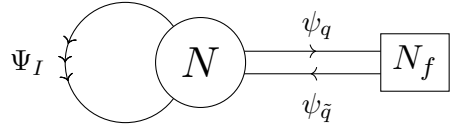

%%%%%
We now proceed to study the SCI of the 3d $\mathcal{N}=2$ holographic SCFT dual to M-theory on AdS$_4\times V^{5,2}/\mathbb{Z}_{N_f}$, which we call the $V^{5,2}$ theory. We refer the reader to \cite{Jafferis:2009th,Hosseini:2016ume,Bobev:2023lkx} for details about the $V^{5,2}$ theory. The corresponding CS-matter theory in the UV is described by the quiver diagram in Fig.~\ref{quiver:V52}.

As with the $N^{0,1,0}$ theory, we first match the TTI conventions used here to the ones in \cite{Bobev:2023lkx}. For the $V^{5,2}$ theory the Bethe potential (\ref{V}) reads
\begin{align}\label{V:V52}
		&\mV_{V^{5,2}}[u;\Delta,\mn]\nonumber\\
		&=\sum_{i=1}^N\pi(2n_{1,i}-\Delta_{m,1})u_{1,i}\nonumber\\
		&\quad+\sum_{I=1}^3\sum_{i,j=1}^N\bigg[\text{Li}_2(e^{\ri(u_{1,i}-u_{1,j}+\pi\Delta_I)})-\fft14(u_{1,i}-u_{1,j})^2-\fft{\pi\Delta_I}{2}(u_{1,i}-u_{1,j})\bigg]\\
		&\quad+N_f\sum_{i=1}^N\bigg[\text{Li}_2(e^{\ri(u_{1,i}+\pi\Delta_q)})-\fft14u_{1,i}^2-\fft{\pi\Delta_q}{2}u_{1,i}+\text{Li}_2(e^{\ri(-u_{1,i}+\pi\Delta_{\tq})})-\fft14u_{1,i}^2+\fft{\pi\Delta_{\tq}}{2}u_{1,i}\bigg]\,.\nonumber
\end{align}
The TTI BA formula (\ref{TTI:3}) for the $V^{5,2}$ theory reads
\begin{equation}
	\begin{split}
		Z_{V^{5,2}}(\Delta,\mn)&=\sum_{\{u_{1,i}\}\in\text{BAE}}\fft{1}{\det\mathbb{B}}\prod_{i=1}^Nx_{1,i}^{\mft_1}\prod_{i\neq j}^N\bigg(1-\fft{x_{1,i}}{x_{1,j}}\bigg)\\
		&\quad\times\prod_{I=1}^3\prod_{i,j=1}^N\bigg(\fft{e^{\ri(u_{1,i}-u_{1,j}+\pi\Delta_I)/2}}{1-e^{\ri(u_{1,i}-u_{1,j}+\pi\Delta_I)}}\bigg)^{1-\mn_I}\\
		&\quad\times\prod_{i=1}^N\bigg(\fft{e^{\ri(u_{1,i}+\pi\Delta_q)/2}}{1-e^{\ri(u_{1,i}+\pi\Delta_q)}}\bigg)^{N(1-\mn_q)}\bigg(\fft{e^{\ri(-u_{1,i}+\pi\Delta_{\tq})/2}}{1-e^{\ri(-u_{1,i}+\pi\Delta_{\tq})}}\bigg)^{N(1-\mn_{\tq})}\,,\label{TTI:V52}
	\end{split}
\end{equation}
where the flavor chemical potentials and magnetic fluxes are constrained as 
\begin{equation}
	\begin{alignedat}{3}
		\sum_{I=1}^3\Delta_I&=\Delta_1+\Delta_2+\Delta_q+\Delta_{\tq}=2\,,&\qquad&\text{and}&\qquad\Delta_3&=\fft23\,,\\
		\sum_{I=1}^3\mn_I&=\mn_1+\mn_2+\mn_q+\mn_{\tq}=2\,,&\qquad&\text{and}&\qquad\mn_3&=\fft23\,.
	\end{alignedat}\label{V52:constraints}
\end{equation}
The superconformal configuration is given by \cite{Bobev:2023lkx}
\begin{equation}
	\Delta_I=\fft23\,,\qquad\Delta_m=0\,,\qquad\mn_I=\fft23\,,\qquad\mft=0\,.\label{V52:special}
\end{equation}
With this at hand, it is straightforward to check that the expressions (\ref{V:V52}) and (\ref{TTI:V52}) match the Bethe potential and the TTI of the $V^{5,2}$ theory presented in \cite{Bobev:2023lkx} (see Appendix \ref{app:Bethe:V52} below for more details). We can therefore employ the numerical BAE solutions constructed in \cite{Bobev:2023lkx}\footnote{These solve the $V^{5,2}$ BAE derived from \eqref{V:V52} for the choice $n_{1,i} =\left\lfloor\fft{N+1}{2}\right\rfloor-i$.}
for various configurations satisfying the constraints (\ref{V52:constraints}). The closed form expression for the $V^{5,2}$ TTI obtained by substituting those numerical solutions in (\ref{TTI:V52}) reads \cite{Bobev:2023lkx}
\begin{equation}
	\begin{split}
		&\Re\log Z^{V^{5,2}}(\Delta,\mn)\\
		&=-\fft{\pi\sqrt{N_f\tilde\Delta_1\tilde\Delta_2\tilde\Delta_3\tilde\Delta_4}}{3}\sum_{a=1}^4\fft{\tilde\mn_a}{\tilde\Delta_a}(\hat N_{N_f,\tDelta})^\fft32\\
		&\quad-\fft{\pi\sqrt{N_f\tilde\Delta_1\tilde\Delta_2\tilde\Delta_3\tilde\Delta_4}}{3}\left(\sum_{I=1}^2(\mathfrak{a}_IN_f+\fft{\mathfrak{b}_I}{N_f})\mn_I+\fft{\tDelta_3-\tDelta_4}{3\tDelta_3^2\tDelta_4^2}\fft{\mft}{N_f^2}\right)(\hat N_{N_f,\tDelta})^\fft12\\
		&\quad-\fft12\log\hat N_{N_f,\tDelta}+\hat f_0(N_f,\tilde\Delta,\tilde\mn)+\hat f_\text{np}(N,N_f,\tDelta,\tmn) \, ,
	\end{split}\label{TTI:V52:result}
\end{equation}
where the shifted $N$ parameter is given by
\begin{equation}
	\hat N_{N_f,\tDelta}=N-\fft{2-\Delta_q-\Delta_{\tq}}{\Delta_3}\fft{N_f}{24}+\fft{N_f}{12}\bigg(\fft{1}{\tDelta_1}+\fft{1}{\tDelta_2}\bigg)+\fft{1}{12N_f}\bigg(\fft{1}{\tDelta_3}+\fft{1}{\tDelta_4}\bigg)\,,\label{V52:details}%
\end{equation}
and we have also defined
\begin{subequations}
	\begin{align}
		\tilde\Delta_a&=\bigg(\Delta_1,\Delta_2,\fft{2-\Delta_q-\Delta_{\tq}}{2}-\fft{\Delta_m}{N_f},\fft{2-\Delta_q-\Delta_{\tq}}{2}+\fft{\Delta_m}{N_f}\bigg)\,,\\
		\tilde\mn_a&=\bigg(\mn_1,\mn_2,\fft{2-\mn_q-\mn_{\tq}}{2}+\fft{\mathfrak t}{N_f},\fft{2-\mn_q-\mn_{\tq}}{2}-\fft{\mathfrak t}{N_f}\bigg)\,,\\
		\mathfrak a_I(\tilde\Delta)&=-\fft{1}{\tilde\Delta_I}\fft{2-(\fft23-\tilde\Delta_I)}{4\tilde\Delta_1\tilde\Delta_2}\,, \\
		\mathfrak b_I(\tilde\Delta)&=-\fft{2}{3\tilde\Delta_1\tilde\Delta_2\tilde\Delta_3\tilde\Delta_4}-\fft{3}{4\tDelta_3\tDelta_4}-\fft{(\tDelta_3-\tDelta_4)^2}{8\tDelta_3^2\tDelta_4^2}\,.
	\end{align}\label{V52:details:2}%
\end{subequations}
The non-perturbative correction in (\ref{TTI:V52:result}) is exponentially suppressed at large $N$, $\hat f_\text{np}(N,N_f,\tDelta,\tmn)\sim\mO(e^{-\sqrt{N}})$, while numerical estimates for the $N$-independent term $\hat f_0$ can be found in \cite{Bobev:2023lkx}. Some of those numerical values for the superconformal flavor magnetic flux configuration, $\tmn_\text{sc}$, are presented below.\footnote{Some of the numerical values of $\hat f_0$ in this table are new and not given explicitly in \cite{Bobev:2023lkx} since only selected numerical data were presented there.}
\begin{center}
	\footnotesize
	\renewcommand*{\arraystretch}{1.2}
	\begin{tabular}{ |c|c||c|c| }
		\hline
		$(N_f,\Delta_1,\Delta_q,\Delta_m)$ & $\hat f_0(N_f,\tDelta,\tmn_\text{sc})$ & $(N_f,\Delta_1,\Delta_q,\Delta_m)$ & $\hat f_0(N_f,\tDelta,\tmn_\text{sc})$ \\
		\hline\hline
		$(1,\fft23,\fft13,0)$ & $-2.7620858097124988759$ & $(3,\fft59,\fft13,\fft{N_f}{9})$ & $-4.8860565481247352522$ \\
		\hline
		$(3,\fft23,\fft13,0)$ & $-4.7624014875151824187$ & $(2,\fft{7}{12},\fft13,\fft{N_f}{15})$ & $-3.3109278391444872740$ \\
		\hline
		$(2,\fft12,\fft13,0)$ & $-3.4523973380968433835$ & $(1,\fft23-\fft{1}{2\pi},\fft16,N_f(\fft23-\fft2\pi))$ & $-2.8401820199427569809$ \\ 
		\hline
	\end{tabular}
\end{center}

Now we proceed to the SCI of the $V^{5,2}$ theory. Similar to the $N^{0,1,0}$ theory case, the only missing component for determining the first two leading terms in the Cardy-like expansion of the $V^{5,2}$ SCI is a closed-form expression for the Bethe potential. Using the numerical BAE solutions $\{u_\star\}$ constructed in \cite{Bobev:2023lkx}, we find that the following analytic expression is in excellent agreement with the numerical data
\begin{equation}
	\Im\mV_{V^{5,2}}[u_\star;\Delta,\mn]=2\pi\left[\fft{\pi\sqrt{N_f\tDelta_1\tDelta_2\tDelta_3\tDelta_4}}{3}\hat N_{N_f,\Delta}^\fft32+\hat g_0(N_f,\Delta)+\hat g_\text{np}(N,N_f,\Delta)\right]\,.\label{mW:Cardy:0:V52:saddle} 
\end{equation}
The non-perturbative correction is exponentially suppressed at large $N$, i.e. $\hat g_\text{np}(N,N_f,\Delta)\sim\mO(e^{-\sqrt{N}})$. 
Below we present numerical values of $\hat g_0$ in selected examples.
\begin{center}
	\footnotesize
	\renewcommand*{\arraystretch}{1.2}
	\begin{tabular}{ |c|c||c|c| }
		\hline
		$(N_f,\Delta_1,\Delta_q,\Delta_m)$ & $\hat g_0(N_f,\Delta)$ & $(N_f,\Delta_1,\Delta_q,\Delta_m)$ & $\hat g_0(N_f,\Delta)$ \\
		\hline\hline
		$(1,\fft23,\fft13,0)$ & $-0.13945567062297404931$ & $(3,\fft59,\fft13,\fft{N_f}{9})$ & $-0.30398703304578607507$ \\
		\hline
		$(3,\fft23,\fft13,0)$ & $-0.29657318363534945350$ & $(2,\fft{7}{12},\fft13,\fft{N_f}{15})$ & $-0.19117259832590889123$ \\
		\hline
		$(2,\fft12,\fft13,0)$ & $-0.20242609578739029964$ & $(1,\fft23-\fft{1}{2\pi},\fft16,N_f(\fft23-\fft2\pi))$ & $-0.14096914088723067753$ \\ 
		\hline
	\end{tabular}
\end{center}
See Appendix \ref{app:Bethe:V52} for numerical data that supports the analytic expression (\ref{mW:Cardy:0:V52:saddle}). \\

Substituting the TTI (\ref{TTI:V52:result}) and the Bethe potential (\ref{mW:Cardy:0:V52:saddle}) into the Cardy-like expansion (\ref{summary:improved}), we obtain the SCI for the $V^{5,2}$ theory in the Cardy-like limit 
\begin{equation}
	\begin{split}
		&\log\mI_{V^{5,2}}(\omega,\Delta,\mn)\\
		&=-\fft{2}{\omega}\left[\fft{\pi\sqrt{N_f\tDelta_1\tDelta_2\tDelta_3\tDelta_4}}{3}\hat N_{N_f,\Delta}^\fft32+\hat g_0(N_f,\Delta)+\hat g_\text{np}(N,N_f,\Delta)\right]\\
		&\quad+\Bigg[-\fft{\pi\sqrt{N_f\tilde\Delta_1\tilde\Delta_2\tilde\Delta_3\tilde\Delta_4}}{3}\sum_{a=1}^4\fft{\tilde\mn_a}{\tilde\Delta_a}(\hat N_{N_f,\tDelta})^\fft32\\
		&\kern3em-\fft{\pi\sqrt{N_f\tilde\Delta_1\tilde\Delta_2\tilde\Delta_3\tilde\Delta_4}}{3}\left(\sum_{I=1}^2(\mathfrak{a}_IN_f+\fft{\mathfrak{b}_I}{N_f})\mn_I+\fft{\tDelta_3-\tDelta_4}{3\tDelta_3^2\tDelta_4^2}\fft{\mft}{N_f^2}\right)(\hat N_{N_f,\tDelta})^\fft12\\
		&\kern3em-\fft12\log\hat N_{N_f,\tDelta}+\hat f_0(N_f,\tilde\Delta,\tilde\mn)+\hat f_\text{np}(N,N_f,\tDelta,\tmn)\Bigg]+\ri\varphi+\mathcal O(\omega)\,.
	\end{split}\label{SCI:V52}
\end{equation}
%

%%%%%
\subsection{$Q^{1,1,1}$ theory}\label{sec:ex:Q111}
\begin{figure}
	\centering
	\begin{tikzpicture}
		\draw[->-=0.47,->-=0.57] (0.57,0.15) -- (2.43,0.15);
		\draw[->-=0.47,->-=0.57] (2.43,-0.15) -- (0.57,-0.15);
		\draw[->-=0.28,->-=0.74] (3,0) arc (0:180:1.5);
		\draw[->-=0.28,->-=0.74] (3,0) arc (0:-180:1.5);
		\node at (0,0) [circle,draw,scale=1.5,fill=white] {$N$};
		\node at (3,0) [circle,draw,scale=1.5,fill=white] {$N$};
		\node at (1.5,1.5) [rectangle,draw,scale=1,fill=white] {$N_f$};
		\node at (1.5,-1.5) [rectangle,draw,scale=1,fill=white] {$N_f$};
		\node at (1.6,0.5) {$\Psi_{1,2}$};
		\node at (1.6,-0.6) {$\Psi_{3,4}$};
		\node at (2.9,1.4) {$\psi_{q_1}$};
		\node at (3,-1.4) {$\psi_{q_2}$};
		\node at (0.1,1.4) {$\psi_{\tq_1}$};
		\node at (0.1,-1.4) {$\psi_{\tq_2}$};
	\end{tikzpicture}
	\caption{Quiver diagram for the $Q^{1,1,1}$ theory}\label{quiver:Q111}
\end{figure}
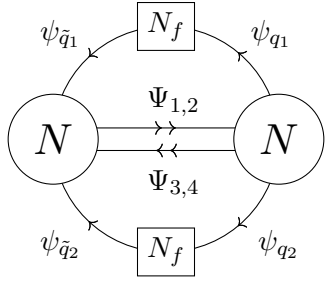

%%%%%
As a third example, we consider the SCI of the 3d $\mathcal{N}=2$ holographic SCFT dual to M-theory on AdS$_4\times Q^{1,1,1}/\mathbb{Z}_{N_f}$. We refer the readers to \cite{Benini:2009qs,Cremonesi:2010ae,Hosseini:2016ume,Bobev:2023lkx} for details about this $Q^{1,1,1}$ theory. The corresponding CS-matter theory in the UV is described by the quiver diagram in Fig.~\ref{quiver:Q111}. 

As in the previous examples we first write down the Bethe potential for the $Q^{1,1,1}$ theory using the general formula (\ref{V}) 
\begin{equation}
	\begin{split}
		&\mV_{Q^{1,1,1}}[u;\Delta,\mn]\\
		&=\sum_{i=1}^N\bigg[\pi(2n_{1,i}-\Delta_{m,1})u_{1,i}+\pi(-2n_{2,i}-\Delta_{m,2})u_{2,i}\bigg]\\
		&\quad+\sum_{a=1}^2\sum_{i,j=1}^N\bigg[\text{Li}_2(e^{\ri(u_{1,i}-u_{2,j}+\pi\Delta_a)})-\fft14(u_{1,i}-u_{2,j})^2-\fft{\pi\Delta_a}{2}(u_{1,i}-u_{2,j})\bigg]\\
		&\quad+\sum_{a=3}^4\sum_{i,j=1}^N\bigg[\text{Li}_2(e^{\ri(u_{2,j}-u_{1,i}+\pi\Delta_a)})-\fft14(u_{2,j}-u_{1,i})^2-\fft{\pi\Delta_a}{2}(u_{2,j}-u_{1,i})\bigg]\\
		&\quad+N_f\sum_{n=1}^2\sum_{i=1}^N\bigg[\text{Li}_2(e^{\ri(-u_{1,i}+\pi\Delta_{\tq_n})})-\fft14u_{1,i}^2+\fft{\pi\Delta_{\tq_n}}{2}u_{1,i}\bigg]\\
		&\quad+N_f\sum_{n=1}^2\sum_{i=1}^N\bigg[\text{Li}_2(e^{\ri(u_{2,i}+\pi\Delta_{q_n})})-\fft14u_{2,i}^2-\fft{\pi\Delta_{q_n}}{2}u_{2,i}\bigg]\,.
	\end{split}\label{V:Q111}
\end{equation}
The general BA formula for the TTI (\ref{TTI:3}) can be written explicitly for the $Q^{1,1,1}$ theory as
\begin{equation}
	\begin{split}
		Z_{Q^{1,1,1}}(\Delta,\mn)&=\sum_{\{u_{r,i}\}\in\text{BAE}}\fft{1}{\det\mathbb{B}}\prod_{r=1}^2\Bigg[\prod_{i=1}^Nx_{r,i}^{\mft_r}\prod_{i\neq j}^N\bigg(1-\fft{x_{r,i}}{x_{r,j}}\bigg)\Bigg]\\
		&\quad\times\prod_{i,j=1}^N\Bigg[\prod_{a=1}^2\bigg(\fft{e^{\ri(u_{1,i}-u_{2,j}+\pi\Delta_a)/2}}{1-e^{\ri(u_{1,i}-u_{2,j}+\pi\Delta_a)}}\bigg)^{1-\mn_a}\times\prod_{a=3}^4\bigg(\fft{e^{\ri(u_{2,j}-u_{1,i}+\pi\Delta_a)/2}}{1-e^{\ri(u_{2,j}-u_{1,i}+\pi\Delta_a)}}\bigg)^{1-\mn_a}\Bigg]\\
		&\quad\times\prod_{n=1}^2\prod_{i=1}^N\bigg(\fft{e^{\ri(-u_{1,i}+\pi\Delta_{\tq_n})/2}}{1-e^{\ri(-u_{1,i}+\pi\Delta_{\tq_n})}}\bigg)^{N_f(1-\mn_{\tq_1})}\bigg(\fft{e^{\ri(u_{2,i}+\pi\Delta_{q_n})/2}}{1-e^{\ri(u_{2,i}+\pi\Delta_{q_n})}}\bigg)^{N_f(1-\mn_{q_n})}\,.\label{TTI:Q111}
	\end{split}
\end{equation}
Here the flavor chemical potentials and magnetic fluxes are constrained as
\begin{equation}
	\begin{alignedat}{3}
		2&=\sum_{a=1}^4\Delta_a&&=\Delta_1+\Delta_{q_1}+\Delta_{\tq_1}&&=\Delta_2+\Delta_{q_2}+\Delta_{\tq_2}\,,\\
		2&=\sum_{a=1}^4\mn_a&&=\mn_1+\mn_{q_1}+\mn_{\tq_1}&&=\mn_2+\mn_{q_2}+\mn_{\tq_2}\,,
	\end{alignedat}\label{Q111:constraints}
\end{equation}
and the superconformal configuration reads
\begin{equation}
	\begin{alignedat}{3}
		\Delta_1&=\Delta_2\,,&\qquad\Delta_3&=\Delta_4\,,&\qquad\Delta_{q_{1,2}}&=\Delta_{\tq_{1,2}}\,,\\
		\mn_1&=\mn_2\,,&\qquad\mn_3&=\mn_4\,,&\qquad\mn_{q_{1,2}}&=\mn_{\tq_{1,2}}\,.\label{Q111:special}
	\end{alignedat}
\end{equation}
In Appendix \ref{app:Bethe:Q111} we show that the expressions (\ref{V:Q111}) and (\ref{TTI:Q111}) match the Bethe potential and the TTI of the $Q^{1,1,1}$ theory presented in \cite{Bobev:2023lkx}. Note that the numerical BAE solutions constructed in this reference solve the BAE for a choice of integers $(n_{1,i},n_{2,i})=(1-i-\left\lfloor N_f/2\right\rfloor+N,i-\left\lfloor N_f/2\right\rfloor)$, and they were obtained for the special configuration (\ref{Q111:special}). At this superconformal point, the BA formula (\ref{TTI:Q111}) yields \cite{Bobev:2023lkx}
\begin{equation}
	\begin{split}
		\Re\log Z_{Q^{1,1,1}}(\Delta_{\text{sc}},\mn_{\text{sc}})&=-\fft{4\pi\sqrt{N_f}}{3\sqrt3}\left((\hat N_{N_f})^\fft32-\bigg(\fft{N_f}{4}+\fft{3}{4N_f}\bigg)(\hat N_{N_f})^\fft12\right)\\
		&\quad-\fft12\log\hat N_{N_f}+\hat f_0(N_f)+\hat{f}_\text{np}(N,N_f) \, ,
	\end{split}\label{TTI:Q111:result}
\end{equation}
where the shifted $N$ parameter is given by
\begin{equation}
	\hat N_{N_f}=N+\fft{N_f}{6}\,.\label{Q111:details}
\end{equation}
As in the previous example we denote the exponentially suppressed correction as $\hat{f}_\text{np}(N,N_f)\sim\mO(e^{-\sqrt{N}})$, and numerical values of $\hat f_0(N_f)$ can be found in \cite{Bobev:2023lkx}. \\

For the SCI of the $Q^{1,1,1}$ theory it suffices to determine the closed form expression of the Bethe potential by using various BAE solutions $\{u_\star\}$ provided in \cite{Bobev:2023lkx}. The analytic expression we find from this numerical data is 
\begin{equation}
	\Im\mV_{Q^{1,1,1}}[u_\star;\Delta_{\text{sc}},\mn]=2\pi\Bigg[\fft{\pi\sqrt{N_f}}{3\sqrt3}(\hat N_{N_f})^\fft32+\hat g_0(N_f)+\hat{g}_\text{np}(N,N_f)\Bigg]\,,\label{mW:Cardy:0:Q111:saddle} 
\end{equation}
at the special configuration (\ref{Q111:special}). The non-perturbative correction is exponentially suppressed at large $N$, $\hat{g}_\text{np}(N,N_f)\sim\mO(e^{-\sqrt{N}})$. Below we present numerical values of $\hat g_0$ together with those of $\hat{f}_0$ given in \cite{Bobev:2023lkx}. 
\begin{center}
	\footnotesize
	\renewcommand*{\arraystretch}{1.2}
	\begin{tabular}{ |c|c|c| }
		\hline
		$N_f$ & $\hat g_0(N_f)$ & $\hat f_0(N_f)$ \\
		\hline\hline
		$1$ & $-0.12179382823357287453$ & $-2.1415723730798296354$ \\
		\hline
		$2$ & $-0.060896914126385874431$ & $-2.0385864384989237526$ \\
		\hline
		$3$ & $0.018581373235204659187$ & $-2.2368141361938090934$ \\
		\hline
		$4$ & $0.12639484451282630333$ & $-2.6005901148883909862$ \\
		\hline
		$5$ & $0.26400260477995552485$ & $-3.1045097958934355205$ \\
		\hline
	\end{tabular}
\end{center}
See Appendix \ref{app:Bethe:Q111} for numerical data that supports the proposed (\ref{mW:Cardy:0:Q111:saddle}). \\

Substituting the TTI (\ref{TTI:Q111:result}) and the Bethe potential (\ref{mW:Cardy:0:Q111:saddle}) into the Cardy-like expansion (\ref{summary:improved}), we obtain the final form of the SCI for the $Q^{1,1,1}$ theory in the Cardy-like limit 
\begin{equation}
	\begin{split}
		\log\mI_{Q^{1,1,1}}(\omega,\Delta_{\text{sc}},\mn_{\text{sc}})\Big|_\text{(\ref{Q111:special})}&=-\fft{2}{\omega}\Bigg[\fft{\pi\sqrt{N_f}}{3\sqrt3}(\hat N_{N_f})^\fft32+\hat g_0(N_f)+\hat{g}_\text{np}(N,N_f)\Bigg]\\
		&\quad+\Bigg[-\fft{4\pi\sqrt{N_f}}{3\sqrt3}\left((\hat N_{N_f})^\fft32-\bigg(\fft{N_f}{4}+\fft{3}{4N_f}\bigg)(\hat N_{N_f})^\fft12\right)\\
		&\kern3em-\fft12\log\hat N_{N_f}+\hat f_0(N_f)+\hat{f}_\text{np}(N,N_f)\Bigg]+\mathcal O(\omega)\,.
	\end{split}\label{SCI:Q111}
\end{equation}
%

%%%%%
\section{Holography }\label{sec:holo}
%%%%%

We now proceed with a discussion on the holographic implications of the explicit expressions for the SCI of the 3d $\mN=2$ SCFTs in Section~\ref{sec:ex} and the index relation (\ref{summary:improved}). 

In most of the discussion below we will focus on the superconformal, or universal, configuration of the fugacities and real masses for the TTI and SCI. As explained in \cite{Azzurli:2017kxo,Bobev:2017uzs,Bobev:2023lkx}, this choice of parameters corresponds to turning on sources and expectation values only for the background fields that couple to the energy-momentum multiplet in the CFT. The holographic dual manifestation of this choice is that the corresponding 4d gravitational backgrounds can be described as solutions of minimal 4d $\mathcal{N}=2$ gauged supergravity.

For this choice of parameters, the field theory results for the TTI and SCI in Sections~\ref{sec:general} and \ref{sec:ex}, along with our previous works \cite{Bobev:2022wem} summarized in Appendix \ref{app:ABJMADHM}, can be succinctly expressed as
\begin{subequations}
\begin{alignat}{2}
	\text{TTI :}&~~-\Re\log Z&&=\pi\alpha\Big((N - B)^\fft32 + C(N - B)^\fft12\Big)\nn\\
	&&&\quad + \fft12\log(N-B) - \hat{f}_0 + \mathcal{O}(e^{-\sqrt{N}})\,,\label{TTI:universal}\\
	\text{SCI :}&\qquad-\log\mI&&=\fft{\pi\alpha}{2\omega}(N-B)^\fft32 + \fft2\omega\hat{g}_0 + \pi\alpha\Big((N - B)^{3/2} + C(N - B)^\fft12\Big) \nn\\
	&&&\quad + \fft12\log(N-B) - \hat{f}_0 + \mathcal{O}(e^{-\sqrt{N}},\omega)\,.\label{SCI:universal}
\end{alignat}\label{index:universal}%
\end{subequations}
The quantities $(\alpha,B,C)$ that determine both indices are presented for various holographic SCFTs in Table~\ref{tab:loc}. As discussed above, the constants $\hat{g}_0$ and $\hat{f}_0$, which are independent of $N$, are not known analytically as functions of the CS level $k$ or the number of fundamental multiplets $N_f$ and assume distinct values for different holographic SCFTs. We note that the quantities $B$ and $C$ for the $V^{5,2}$ and $Q^{1,1,1}$ were only partially determined in our previous work \cite{Bobev:2023lkx} where the combination $B+C$ was calculated. As we discuss below, the relation between the SCI and TTI we derived above allows to determine $B$ and $C$ separately.
\begin{table}
	\centering
	\renewcommand*{\arraystretch}{1.5}
	\begin{tabular}{|c|c|c|c|c|}
		\hline
		$Y_7$ & $\mathcal{N}$ & $\alpha$ & $B$ & $C$ \\
		\hline\hline
		$S^7/\mathbb{Z}_k$ (free) & $6$ & $\frac{\sqrt{2k}}{3}$ & $\frac{k}{24} - \frac{2}{3k}$ & $-\frac{3}{k}$ \\
		\hline
		$S^7/\mathbb{Z}_{N_f}$ (fixed points) & $4$ & $\frac{\sqrt{2N_f}}{3}$ & $-\frac{7N_f}{24} - \frac{1}{3N_f}$ & $-\frac{N_f}{2} -\frac{5}{2N_f}$ \\
		\hline
		$N^{0,1,0}/\mathbb{Z}_k$ & $3$ & $\frac{4\sqrt{k}}{3\sqrt{3}}$ & $-\frac{5k}{48} - \frac{1}{3k}$ & $-\frac{k}{4} - \frac{5}{4k}$ \\
		\hline
		$V^{5,2}/\mathbb{Z}_{N_f}$ & $2$ & $\frac{16\sqrt{N_f}}{27}$ & $-\frac{N_f}{6} - \frac{1}{4N_f}$ & $-\frac{9N_f}{16} - \frac{27}{16N_f}$ \\
		\hline
		$Q^{1,1,1}/\mathbb{Z}_{N_f}$ & $2$ & $\frac{4\sqrt{N_f}}{3\sqrt{3}}$ & $-\frac{N_f}{6}$ & $-\frac{N_f}{4} - \frac{3}{4N_f}$ \\
		\hline
	\end{tabular}
	\caption{The quantities $(\alpha,B,C)$ that determine the TTI and SCI for different 3d $\mathcal{N}\geq 2$ holographic SCFTs. The first two entries are mirror dual to each other for $k=N_f=1$ and correspond to the ABJM and ADHM theories, respectively. Also, we have emphasized that the orbifold action for the ABJM theory has no fixed points contrary to that for the ADHM model. For the third entry, we have imposed $r=k$ as described in Section~\ref{sec:ex:N010}. \label{tab:loc}}
\end{table}
% 

%%%%%
\subsection{Leading term at large $N$}\label{sec:holo:largeN}
%%%%% 
The SCI of the 3d $\mN=2$ holographic SCFTs of interest is holographically dual to the Euclidean path integral of M-theory around the 11d background obtained by uplifting the 4d Euclidean supersymmetric Kerr-Newman (KN) AdS black hole solution of $\mN=2$ minimal gauged supergravity \cite{Kostelecky:1995ei,Caldarelli:1998hg}. This uplift is guaranteed to exist and is given explicitly by  the consistent truncation of 11d supergravity on 7d Sasaki-Einstein (SE) manifolds \cite{Gauntlett:2007ma}. This holographic duality is discussed extensively in the literature, see for example, \cite{Cassani:2019mms,Bobev:2019zmz,Choi:2019dfu,Bobev:2021oku,Bobev:2022wem}), and can be succinctly expressed as\footnote{The subscript ``$\text{f}$'' in the product symbol indicates that $Y_7$ is fibered over the 4d non-compact manifold.}
\begin{equation}
	Z_\text{M-theory}\Big|_{\text{KN EAdS}_4\times_\text{f}\, Y_7}=\mI_{\text{SCFT}_3}\,,\label{Z=I}
\end{equation}
where the 3d SCFT is specified by the internal manifold $Y_7$, see Table~\ref{tab:loc} for the examples we consider in this work. 

In the semi-classical limit where the 4d Newton constant, $G_N$, is much smaller than the square of the EAdS$_4$ radius, $L^2$, the Euclidean path integral dual to the SCI can be approximated by the on-shell action of the two-derivative $\mN=2$ minimal gauged supergravity Euclidean supersymmetric  KN AdS black hole solution \cite{Cassani:2019mms,Bobev:2019zmz,Bobev:2021oku}
\begin{equation}
\begin{split}
	-\log Z_\text{M-theory}\Big|_{\text{KN EAdS}_4\times_\text{f}\, Y_7}&=S^{(2\partial)}_{\mN=2~\text{sugra}}\Big|_{\text{KN EAdS}_4}+o(L^2/G_N)\\
	&=\fft{\pi L^2}{2G_N}\fft{(\omega+1)^2}{2\omega}+o(L^2/G_N)\,.
\end{split}\label{Z:semi-cl}
\end{equation}
Here $\omega$ is the angular velocity of the black hole solution which via supersymmetry also determines the electric charge, see \cite{Cassani:2019mms,Bobev:2019zmz} for more details on the gravitational background and the evaluation of this on-shell action.

We can now use the AdS$_4$/CFT$_3$ dictionary to map the 4d gravitational parameters to the number $N$ of M2-branes at the tip of the cone over the internal manifold $Y_7$, see \cite{Gauntlett:2007ma,Marino:2011nm}. To leading order in the large $N$ limit one finds
\begin{equation}
	\fft{L^2}{2G_N}=\sqrt{\fft{2\pi^4}{27\text{vol}[Y_7]}}\,N^\fft32+o(N^\fft32)\,,\label{AdS4CFT3}
\end{equation}
where $\text{vol}[Y_7]$ is the volume of $Y_7$. Using this we can express the Euclidean path integral (\ref{Z:semi-cl}) at leading order in the large $N$ limit as
\begin{equation}
	-\log Z_\text{M-theory}\Big|_{\text{KN EAdS}_4\times_\text{f}\, Y_7}=\sqrt{\fft{2\pi^6}{27\text{vol}[Y_7]}}\,\fft{(\omega+1)^2}{2\omega}\,N^\fft32+o(N^\fft32)\,.\label{Z:largeN}
\end{equation}
For all Sasaki-Einstein orbifolds listed in Table~\ref{tab:loc} one can show that $\alpha=\sqrt{\fft{2\pi^4}{27\text{vol}[Y_7]}}$, see Appendix E of \cite{Bobev:2023lkx} for more details on this calculation. This in turn shows that in the large $N$ limit the SCI in (\ref{SCI:universal}) reads
\begin{equation}
	-\log\mI_{\text{SCFT}_3}=\fft{\pi\alpha}{2}\bigg(\fft1\omega+2+\mO(\omega)\bigg)N^\fft32+o(N^\fft32)\label{SCI:largeN}
\end{equation}
and indeed precisely agrees with the gravitational on-shell action. A notable difference between the field theory and gravitational results is that the supergravity on-shell action is evaluated for any finite value of $\omega$ while in the field theory calculation of the SCI we only keep the leading two terms in the Cardy-like limit. It will of course be very interesting to establish the holographic duality for general values of $\omega$, see \cite{GonzalezLezcano:2022hcf,BenettiGenolini:2023rkq} for some recent work along this direction.\\

An important holographic application of the SCI is that it provides a microscopic description of the entropy of the dual  4d supersymmetric KN AdS black hole. To obtain this entropy to leading order in the large $N$ limit one needs to perform a suitable Legendre transform. This was done in various contexts in the literature, see for instance \cite{Choi:2018fdc,Hristov:2019mqp,Bobev:2019zmz,Cassani:2019mms}, so we will be brief. One first defines the entropy function\footnote{In the discussion below we assume that the expression for the SCI at finite $\omega$ agrees with the leading order holographic result in \eqref{Z:largeN}.} 
\begin{equation}
	\mathcal{S}(\omega,\varphi,J,Q,\lambda)=-2\pi\alpha\fft{\varphi^2}{\omega}N^\fft32-2\pi\ri\Big(\omega J+\varphi LQ+\lambda(2\varphi-1-\omega)\Big)\,,\label{S-function}
\end{equation}
which is then extremized with respect to the chemical potentials $(\omega,\varphi)$ and the Lagrange multiplier $\lambda$ to obtain the Bekenstein-Hawking entropy of the KN black hole. The information about the microscopic details of the dual SCFT, or alternatively the particular embedding of the 4d black hole solution in 11d supergravity, is encoded in the parameter $\alpha$ which takes the values given in Table~\ref{tab:loc} for the examples studied in this work. Our results for the large $N$ SCI therefore constitute a non-trivial microscopic counting of the entropy of the 4d supersymmetric KN AdS black hole for all these setups arising from the low-energy dynamics of M2-branes.

%%%%%
\subsection{Subleading terms at large $N$}
\label{sec:holo:beyond}
%%%%%

We now turn our attention to the holographic implication of the indices in (\ref{index:universal}) beyond the leading $N^{3/2}$ order in the large $N$ limit with a focus on the 4-derivative corrections to 4d $\mN=2$ minimal gauged supergravity recently analyzed in \cite{Bobev:2020egg,Bobev:2021oku}. The key statement of \cite{Bobev:2020egg,Bobev:2021oku} is that the four-derivative corrections to the Euclidean 4d $\mN=2$ minimal gauged supergravity action are characterized by two dimensionless constant coefficients $\lambda_{1,2}$. Moreover, any classical solution of the 2-derivative equations of motion automatically solves the 4-derivative ones and one can explicitly calculate its on-shell action. The result for the 4-derivative regularized on-shell action $I_{4\partial}$ for any, not necessarily supersymmetric, solution $\mathbb{S}$ is given by
\begin{equation}
	I_{4\partial}(\mathbb{S})=\bigg[1+\fft{64\pi G_N}{L^2}(\lambda_2-\lambda_1)\bigg]\fft{\pi L^2}{2G_N}\mF(\mathbb{S})+32\pi^2\lambda_1\,\chi(\mathbb{S})\,,\label{I4:1}
\end{equation}
where $\chi$ is the regularized Euler characteristic of the 4d Euclidean manifold and $\mF$ determines the 2-derivative regularized on-shell action of the solution $\mathbb{S}$ in a normalization in which the regularized on-shell action of empty Euclidean AdS$_4$ has $\mathcal{F}=1$, see \cite{Bobev:2021oku} for more details. \\

To compare the 4-derivative regularized on-shell action (\ref{I4:1}) with a dual SCFT partition function, we should extend the AdS/CFT dictionary (\ref{AdS4CFT3}) to the subleading order in the large $N$ limit as
\begin{equation}
	\fft{L^2}{2G_N} = A\,N^\fft32 + a\,N^\fft12 + o(N^\fft12)\, , \qquad 32\pi \lambda_i = v_i\,N^\fft12 + o(N^\fft12) \, ,
\end{equation}
where $(A,a,v_1,v_2)$ are real constants that do not scale with $N$. This can then be used to rewrite (\ref{I4:1}) as
\begin{equation}
	I_{4\partial}(\mathbb{S}) = \pi\mathcal{F}(\mathbb{S})\Big(A\,N^\fft32 + (a + v_2)\,N^\fft12\Big) - \pi\Big(\mathcal{F}(\mathbb{S}) - \chi(\mathbb{S})\Big) \,v_1\,N^\fft12 + o(N^\fft12) \,. \label{I4:2}
\end{equation}
Note that the parameters $(A,a,v_1,v_2)$ encode the details of the specific SCFT of interest while the parameters $(\mF,\chi)$ are specified by the 4d supergravity solution dual to the particular partition function at hand. \\

The 4d backgrounds holographically dual to the TTI and the SCI are given by the supersymmetric Euclidean Reissner-Nordstr\"om (RN) AdS black hole and the supersymmetric Euclidean KN AdS black hole, respectively. The quantities $(\mF,\chi)$ for these backgrounds are presented in \cite{Bobev:2020egg,Bobev:2021oku} and read
\begin{equation}
\begin{split}
	\text{RN :}&\quad \{\mF(\mathbb{S}_\text{RN}),\chi(\mathbb{S}_\text{RN})\}=\{1,2\}\,,\\
	\text{KN :}&\quad\{\mF(\mathbb{S}_\text{KN}),\chi(\mathbb{S}_\text{KN})\}=\Big\{\fft{(\omega+1)^2}{2\omega},2\Big\}\,.
\end{split}\label{Fchi}
\end{equation}
Using these gravitational parameters and the holographic relations
\begin{equation}
\begin{split}
	I_{4\partial}(\mathbb{S}_\text{RN})&=-\log Z+o(N^\fft12)\,,\\
	I_{4\partial}(\mathbb{S}_\text{KN})&=-\log\mI+o(N^\fft12)\,,
\end{split}
\end{equation}
we can compare the 4-derivative regularized on-shell action (\ref{I4:2}) with the dual SCFT indices\footnote{When taking the logarithm of the two indices, the imaginary part is defined modulo $2\pi\ri\mathbb{Z}$ and therefore can be absorbed in the subleading corrections of order $o(N^\fft12)$ in the large $N$ limit.} (\ref{index:universal}). This in turn allows us to determine the supergravity coefficients $(A,a,v_1,v_2)$ in terms of the field theory data in Table~\ref{tab:loc} to find
\begin{equation}
	A=\alpha\,,\qquad a+v_2=\fft{\alpha}{2}(C-3B)\,,\qquad v_1=\fft{\alpha}{2}C\,.\label{Aav}
\end{equation}
We note that to determine these relations between supergravity and field theory quantities we only need to use the large $N$ expansion of the TTI (\ref{TTI:universal}) together with the SCI relation in \eqref{summary:improved}. Substituting (\ref{Aav}) back into the 4-derivative regularized on-shell action (\ref{I4:2}), we find
\begin{equation}\label{I4:3}
\begin{split}
	I_{4\partial}(\mathbb{S}) &= \pi\alpha\bigg[\mathcal{F}(\mathbb{S})N^{3/2}  +\fft{C\chi(\mathbb{S})-3B\mathcal{F}(\mathbb{S})}{2}N^{1/2}\bigg] + o(N^\fft12)  \\
	&=-\log\mZ_{\partial\mathbb{S}}+o(N^\fft12)\,,
\end{split}
\end{equation}
where $\mZ_{\partial\mathbb{S}}$ denotes the general 3d partition function on the conformal boundary of $\mathbb{S}$. For instance, in this general notation, the TTI and the SCI read $\mZ_{\partial\mathbb{S}_\text{RN}}=Z$ and $\mZ_{\partial\mathbb{S}_\text{KN}}=\mI$, respectively. \\

The expression in \eqref{I4:3} has important implications. It allows for the calculation of the two leading order $N^\fft32$ and $N^\fft12$ terms in the large $N$ expansion of \emph{any} 3d partition function for the 3d $\mN=2$ SCFTs arising from $N$ M2-branes listed in Table~\ref{tab:loc}. All we need to know are the quantities $\mathcal{F}$ and $\chi$ for the 4d Euclidean background corresponding to the given 3d manifold on which the SCFT is placed on. As a first consistency check we note that the SCI for the theories listed in Table~\ref{tab:loc} agrees with \eqref{I4:3} after using the second line in \eqref{Fchi}. This is of course expected and compatible with the field theory relation between the SCI and the TTI in \eqref{summary:improved}. An additional stronger consistency check of the relation (\ref{I4:3}), and a precision test of holography, is provided by considering the $S^3$ partition function of the ABJM, ADHM, and $N^{0,1,0}$ theories. For these models, as summarized in \cite{Bobev:2022eus,Bobev:2023lkx}, the round $S^3$ partition function in the absence of sources can be written in terms of an Airy function. Using the large $N$ expansion of the Airy function, and the fact that for the EAdS$_4$ background with $S^3$ boundary we have $(\mF,\chi)=(1,1)$, one indeed finds that (\ref{I4:3}) is obeyed for these three SCFTs.\footnote{For the leading $N^\fft{3}{2}$ order, \cite{Choi:2019dfu} studied not only the SCI and TTI but also the $S^3$ partition function for general holographic 3d $\mN=2$ SCFTs and observed that (\ref{I4:3}) is indeed obeyed to that order. This is also consistent with the large $N$ universality emphasized in \cite{Bobev:2017uzs}.}

For the theories and path integrals for which large $N$ supersymmetric localization results are not yet available the expression in \eqref{I4:3} provides a valuable supergravity prediction for the 3d SCFT. For instance, for the $V^{5,2}$ and $Q^{1,1,1}$ theories there are no supersymmetric localization results available for the $N^\fft12$ terms of the squashed $S^3$ free energy. Using \eqref{I4:3}  and the fact that $(\mF,\chi)=(\frac{1}{4}(b+1/b)^2,1)$ for the supersymmetric $\rm{U}(1)\times \rm{U}(1)$ squashed $S^3$, see \cite{Bobev:2020egg}, we find the following holographic prediction
	\begin{equation}
		-\log\mZ_{S^3_b}=\pi\alpha\bigg[\frac{1}{4}\left(b+\frac{1}{b}\right)^2N^\fft32+\left(\fft{C}{2}-\frac{3B}{8}\left(b+\frac{1}{b}\right)^2\right)N^\fft12\bigg]+o(N^\fft12)\,,\label{prediction}
	\end{equation}
where the coefficients $(\alpha,B,C)$ for the $V^{5,2}$ and $Q^{1,1,1}$ models can be read off from Table~\ref{tab:loc}. Confirming the prediction (\ref{prediction}) at the $N^\fft12$ order for the $V^{5,2}$ and $Q^{1,1,1}$ theories from supersymmetric localization would be of great interest. Indeed, it was recently shown in \cite{Hong:2024} that (\ref{prediction}) is compatible with a saddle point analysis of the $S^3$ partition function of these models in the IIA limit. In addition, (\ref{prediction}) agrees well with the Airy conjecture for this partition function discussed in~\cite{AiryTale}.

Another important application of the results above is in the holographic calculation of thermal observables in 3d $\mN=2$ holographic SCFTs beyond the strict large $N$ limit. As recently demonstrated in \cite{Bobev:2023ggk} the result in (\ref{I4:3}) can be used to calculate the thermal free energy on $S^1\times \mathbb{R}^2$ as well as some coefficients in the thermal effective action on $S^1\times S^2$ for theories arising from M2-branes to order $N^\fft12$. The results of \cite{Bobev:2023ggk} can be directly applied, in conjunction with (\ref{I4:3}), in order to compute these observables also for all theories in Table~\ref{tab:loc}.

Finally, we make a few comments on the $\log N$ terms in the large $N$ expansion of the TTI and SCI in \eqref{index:universal}. As shown in \cite{Bobev:2023dwx}, see also \cite{Hristov:2021zai}, these logarithmic corrections to the free energies of holographic SCFTs are universal and do not depend on continuous parameters. The large $N$ expansion of the SCI and TTI indeed confirms this general result. Moreover, based on the results in \cite{Bobev:2023dwx}, we conclude that the logarithm of \emph{any} path integral on a compact Euclidean manifold $\mathcal{M}$ for the 3d SCFTs in Table~\ref{tab:loc} (in the presence of arbitrary sources) contains the following universal logarithmic term
\begin{equation}
-\log Z_{\mathcal{M}} \supset \frac{\chi}{4}\log N\,,
\end{equation}
where, as above, $\chi$ is the regularized Euler number of the 4d Euclidean manifold, with $\mathcal{M}$ as its boundary.

%%%%%
\section{Discussion}\label{sec:discussion}
%%%%%

The main result of our work is to establish the relation in \eqref{summary} between the first two leading terms in the Cardy-like expansion of the SCI and the Bethe potential and TTI for a broad class of 3d $\mathcal{N}=2$ SCFTs. We then showed how to utilize this result in the context of 3d $\mN=2$ holographic SCFTs arising from $N$ M2-branes probing certain Calabi-Yau 4-fold singularities. Combining the relation in \eqref{summary} with the exact TTI results of \cite{Bobev:2022jte,Bobev:2022eus,Bobev:2023lkx}, we derived the all order $1/N$-perturbative expansions for the first two leading terms in the Cardy-like expansion of the SCI for a number of such holographic SCFTs. We also discussed how our results can be used in the context of precision holography by combining them with recent advances in our understanding of higher-derivative and logarithmic corrections to 4d gauged supergravity. Below we briefly discuss several open problems and possible generalizations of our work.

One of the most conceptually straightforward, yet technically challenging, questions that remain open is the computation of the SCI for general finite values of the parameter $\omega$. Evaluating the SCI localization formula (\ref{SCI:3}) with a finite $\omega$ is in general very complicated and, even when employing the saddle point approximation for sufficiently small $\omega$, keeping track of subleading corrections beyond the $\omega^0$ order becomes highly non-trivial, see \cite{Bobev:2022wem}. The divergent series expansion of the $\infty$-Pochhammer symbol discussed in Appendix \ref{app:poch} poses another complication in analyzing the SCI perturbatively in $\omega$. A promising starting point to address these difficulties could be to focus on the SCI of holographic SCFTs in the large $N$ limit, see \cite{GonzalezLezcano:2022hcf,BenettiGenolini:2023rkq} for recent attempts in this direction. However, even within this context, the simple $\fft{(\omega+1)^2}{2\omega}$ $\omega$-dependence of the $N^{\frac{3}{2}}$ term in the large $N$ expansion, as predicted by the holographic dual on-shell supergravity action in (\ref{Z:semi-cl}), has not yet been fully reproduced from the field theory perspective and remains an important open problem.

Our work can be generalized by applying the general index relation (\ref{summary:improved}) to different classes of holographic SCFTs beyond the examples arising from M2-branes studied here. Several classes of prime examples for future exploration are theories arising on the worldvolume of D2-branes in massive IIA string theory, or those associated to M5-branes wrapping hyperbolic 3-manifolds, or SCFTs obtained by wrapping D4-branes on a Riemann surface \cite{Seiberg:1996bd,Bah:2018lyv}. For example, for the D2-brane models in massive IIA, the leading $N^{\frac{5}{3}}$ behavior of the Cardy-like limit of the SCI and the TTI were studied in \cite{Choi:2019dfu} and our results above provide the necessary ingredients to extend this beyond the leading order in the large $N$ limit. In pursuing this, the first step would be to evaluate the large $N$ limit of the TTI and subsequently leverage the corresponding Bethe potential and TTI results to derive the small $\omega$ expansion of the SCI using the index relation (\ref{summary:improved}).

The exact results for the first two leading terms of the SCI of $\mN=2$ holographic SCFTs in the Cardy-like expansion presented above can be used to explore not only the perturbative expansion of the holographically dual string/M-theory path integrals but also the associated non-perturbative corrections. In particular, the leading exponentially suppressed term in the large $N$ expansion of the SCI should capture the instanton contributions to the dual string/M-theory path integral. For the $S^3$ partition functions these non-perturbative corrections at large $N$ were studied for the ABJM theory in \cite{Drukker:2010nc,Drukker:2011zy,Hatsuda:2012dt,Hatsuda:2013gj}, as well as in the holographically dual string/M-theory background more recently in \cite{Gautason:2023igo,Beccaria:2023ujc}. It is reasonable to expect that a similar holographic comparison can be performed for the 3d $\mN=2$ SCI using the non-perturbative results for the TTI presented in \cite{Bobev:2022eus,Bobev:2023lkx} together with the index relation (\ref{summary:improved}).

In Section~\ref{sec:holo}, we explored the holographic dual backgrounds to the so-called universal SCI and TTI in \eqref{index:universal}. The field theory results we derived in Section~\ref{sec:ex} are however more general since they allow for the presence of general fugacities for flavor symmetries and real mass parameters. Introducing these additional parameters in supergravity should amount to finding more general supersymmetric asymptotically AdS$_4$ Kerr-Newman and Reissner-Nordstr\"om black hole solutions of 11d supergravity. It will be very interesting to construct and study these solutions explicitly since they will provide a fertile ground for additional precision tests of the holographic duality.

The relation between the Cardy-like limit of the SCI and the TTI in \eqref{summary:improved} can be viewed as a consequence of a relation between the TTI and the ``disk index'' $\mathcal{I}_{S^1\times_\omega D^2}$, i.e. the partition function of 3d $\mathcal{N}=2$ SCFTs on $S^1\times_{\omega}D^2$. As pointed out in \cite{Choi:2019dfu}, the leading order term in the small $\omega$ expansion of the logarithm of the disk index is controlled by the Bethe potential. Our work above implies the more general relation valid for any 3d $\mathcal{N}=2$ SCFT\footnote{The detailed derivation of the relation in \eqref{eq:IS1D2TTI} follows directly the approach of \cite{Choi:2019dfu} combined with the observations in Section~\ref{sec:general} above.}
\begin{equation}\label{eq:IS1D2TTI}
\log \mathcal{I}_{S^1\times_\omega D^2} (\omega, \Delta, \mn) = -\frac{1}{2\pi\ri \omega} \mathcal{V}[u^*;\Delta,\mn]+\frac{1}{2} \log Z(\Delta,\mn) + \frac{1}{2} r_G \log(-\ri \omega) + \mathcal{O}(\omega)\,,
\end{equation}
where $r_G$ denotes the rank of the gauge group, which equals to $pN$ for the theories studied in this paper. \eqref{eq:IS1D2TTI} relates the $\omega^0$ term of $\log \mathcal{I}_{S^1\times_\omega D^2}$ to the TTI, i.e. to the $S^1\times S^2$ partially twisted partition function $\log Z(\Delta,\mn)$. Since the SCI can be thought of, in some sense, as gluing two partition functions on $S^1\times_{\omega}D^2$, the relation between the SCI and TTI in \eqref{summary:improved} could be viewed as a consequence of \eqref{eq:IS1D2TTI}. 

One can also contemplate studying the Euclidean path integrals of 3d $\mathcal{N}=2$ SCFTs on more general compact manifolds. As noted in \cite{Closset:2017zgf,Closset:2018ghr,Closset:2019hyt}, the supersymmetric partition function of 3d $\mN=2$ theories on any Seifert manifold exhibits a close relationship with the TTI and can be expressed in terms of appropriate fibering and handle-gluing operators summed over the Bethe vacua. It will be very interesting to leverage the perturbatively exact TTI results of \cite{Bobev:2022eus,Bobev:2023lkx}, together with the SCI and disk index relations presented above, to understand better the structure of these supersymmetric path integrals and their holographic implications.

More generally, our work could be viewed as a consequence of the factorization properties of supersymmetric partition functions of 3d $\mathcal{N}=2$ theories on Euclidean manifolds \cite{Dimofte:2011py,Beem:2012mb,Hwang:2012jh}. Understanding better the consequences of this type of factorization for large $N$ holographic theories and clarifying its implication for the dual string/M-theory path integrals is an important open problem, see \cite{Choi:2019dfu,Hosseini:2019iad,Hosseini:2021mnn,Hristov:2021qsw,Hristov:2022lcw,Hristov:2024cgj} for some developments in this direction. We hope that our work provides a useful stepping stone in uncovering this structure and studying its consequences for string and M-theory.

%%%%%%%%%%%%%%%%%%%%%%
\section*{Acknowledgments}
%%%%%%%%%%%%%%%%%%%%%%

We are grateful to Arash Arabi Ardehali, Davide Cassani, Pieter-Jan De Smet, Dongmin Gang, Seppe Geukens, Kiril Hristov, Chiung Hwang, Seok Kim, Zohar Komargodski, Silviu Pufu, and Xuao Zhang for valuable discussions. NB is supported in part by FWO projects G003523N, G094523N, and G0E2723N, as well as  by the Odysseus grant G0F9516N from the FWO. SC is supported in part by a KIAS Individual Grant PG081602 at Korea Institute for Advanced Study. SC is grateful to KU Leuven for kind hospitality during part of this project. JH is supported by the Sogang University Research Grant of 202410008.01, the Basic Science Research Program of the National Research Foundation of Korea (NRF) funded by the Ministry of Education through the Center for Quantum Spacetime (CQUeST) with grant number NRF-2020R1A6A1A03047877, and the Fonds Wetenschappelijk Onderzoek--Vlaanderen (FWO) Junior Postdoctoral Fellowship with grant number 1203024N. JH is grateful to KIAS and Seoul National University for the warm hospitality during parts of this project. VR is partly supported by a Visibilit\'e Scientifique Junior Fellowship from LabEx LMH and is grateful to the CCPP at New York University for hospitality during part of this project.

\appendix

%%%%%
\section{$\infty$-Pochhammer symbol}\label{app:poch}
%%%%%
Here we briefly summarize the definition and properties of the $\infty$-Pochhammer symbol. It is defined within the unit disk as
\begin{equation}
	(a;q)_\infty=\prod_{n=0}^\infty(1-aq^n)\,,\qquad(|q|<1)\,,\label{eq:poch}
\end{equation}
and can be extended to $|q|>1$, see Appendix A of \cite{Choi:2019dfu} for more details. The $\infty$-Pochhammer symbol satisfies the identity
\begin{equation}
	(-x)^{\fft{\mm}{2}}\fft{(xq^{1+\mm};q^2)_\infty}{(x^{-1}q^{1+\mm};q^2)_\infty}=(-x)^{-\fft{\mm}{2}}\fft{(xq^{1-\mm};q^2)_\infty}{(x^{-1}q^{1-\mm};q^2)_\infty}\,,\qquad (\mm\in\mathbb{Z})\,,\label{poch:inverse}
\end{equation}
and has the following asymptotic expansion ($q=e^{\ri\pi\omega}$)
\begin{equation}
	\begin{split}
		\lim_{\omega\to \ri0^+}(aq^x;q^2)_\infty&=\exp[-\fft{\ri}{2\pi\omega}\text{Li}_2(aq^{x-1})](1+\mathcal O(\omega))\\
		&=\exp[-\fft{\ri}{2\pi\omega}\text{Li}_2(a)+\fft{x-1}{2}\text{Li}_1(a)](1+\mathcal O(\omega))\,, ~~(a\in\mathbb C,~a\notin[1,\infty))
	\end{split}\label{poch:asymp}
\end{equation}
in terms of the polylogarithm functions. We comment further on the expansion of the $\infty$-Pochhammer symbol beyond the $\omega^0$ order below. Since the asymptotic expansion in (\ref{poch:asymp}) involves the polylogarithm functions, we also provide the inversion formula
\begin{equation}
	\text{Li}_n(e^{\ri x})+(-1)^n\text{Li}_n(e^{-\ri x})=-\fft{(2\pi\ri)^n}{n!}B_n(\fft{x}{2\pi})~~~\begin{cases}
		0\leq\Re[x]<2\pi~\&~\Im[x]\geq0 \\
		0<\Re[x]\leq2\pi~\&~\Im[x]<0
	\end{cases}\,,\label{polylog:inversion}
\end{equation}
which is useful to compare the generic Bethe potential (\ref{V}) with the known expressions for special cases \cite{Bobev:2023lkx}. \\

To go beyond the $\omega^0$ order in the asymptotic expansion (\ref{poch:asymp}), first we assume $|a|<1$ and $|q|<1$. Then one can expand the $\infty$-Pochhammer symbol as
\begin{equation}
\begin{split}
	\log(aq^x,q^2)_\infty&=\sum_{n=0}^\infty\log(1-aq^{2n+x})\\
	&=-\sum_{n=0}^\infty\sum_{r=1}^\infty\fft1r(aq^{2n+x})^r\\
	&=\sum_{r=1}^\infty\fft1ra^r\fft{q^{xr}}{q^{2r}-1}\\
	&=\sum_{r=1}^{\lfloor1/|\omega|\rfloor}\fft1ra^r\sum_{k=0}^\infty B_k(\fft{x}{2})\fft{(2r\ri\pi \omega)^{k-1}}{k!}-\sum_{r=\lfloor1/|\omega|\rfloor+1}^\infty\fft1r(aq^x)^r\fft{1}{1-q^{2r}}\\
	&\neq\sum_{k=0}^\infty B_k(\fft{x}{2})\text{Li}_{2-k}(a)\fft{(2\ri\pi \omega)^{k-1}}{k!}\,,
\end{split}\label{poch:exp}
\end{equation}
where we have used the generating function of Bernoulli polynomials 
\begin{equation}
	\fft{te^{xt}}{e^t-1}=\sum_{k=0}^\infty B_k(x)\fft{t^k}{k!}\,,\quad~(|t|<2\pi)\,,\label{Bernoulli}
\end{equation}
in the 4th line of \eqref{poch:exp}. It is worth nothing a subtle feature. The formula (\ref{Bernoulli}) cannot be applied to the 3rd line of (\ref{poch:exp}) for $r>\lfloor1/|\omega|\rfloor$ due to the finite radius of convergence and therefore the last line of (\ref{poch:exp}) is in general not a valid asymptotic expansion for the $\infty$-Pochhammer symbol. In fact the last line of (\ref{poch:exp}) is a divergent series, see Appendix A of \cite{GonzalezLezcano:2022hcf} for example. Hence in the main text we proceed with the the asymptotic expansion (\ref{poch:asymp}) based on \cite{Fredenhagen:2004cj} and Appendix A of \cite{Choi:2019zpz}.

%%%%%
\section{Phases in the localization of the SCI }\label{app:SCI-phase}
%%%%%
In this Appendix we derive the SCI localization formula (\ref{SCI:1}) based on \cite{Aharony:2013dha,Aharony:2013kma}, see also \cite{Garozzo:2019ejm} for a nice summary of the conventions. 

The localization formula for the $\mN=2$ SCI (\ref{SCI:tr}) is given by \cite{Aharony:2013dha,Aharony:2013kma,Garozzo:2019ejm}
\begin{equation}
	\begin{split}
		\mI(q,\boldsymbol{\xi})&=\fft{1}{|\mW|}\sum_{\mm_1,\cdots,\mm_{r_G}\in\mathbb Z}\oint\Bigg[\prod_{\ell=1}^{r_G}\fft{dz_\ell}{2\pi \ri z_\ell}(\underline{(-1)^{\mm_\ell}}z_\ell)^{\sum_{n=1}^{r_G}k_{\ell n}\mm_n}\Bigg]\times\prod_{x}\xi_x^{\sum_{\ell=1}^{r_G}k_{x\ell}\mm_{\ell}}\\
		&\quad\times\prod_{\alpha\in\mathfrak{g}}q^{-\fft12|\alpha(\mm)|}\left(1-e^{\ri\alpha(h)}q^{|\alpha(\mm)|}\right)\\
		&\quad\times\prod_{\Psi}\prod_{\rho_\Psi}\left(q^{1-R(\Psi)}\underline{(-1)^{-\rho_\Psi(\mm)}}e^{-\ri\rho_\Psi(h)}\prod_{x}\xi_x^{-f_x(\Psi)}\right)^{\fft12|\rho_\Psi(\mm)|}\\
		&\kern5em\times\fft{(e^{-\ri\rho_\Psi(h)}\prod_{x}\xi_x^{-f_x(\Psi)}q^{2-R(\Psi)+|\rho_\Psi(\mm)|};q^2)_\infty}{(e^{\ri\rho_\Psi(h)}\prod_{x}\xi_x^{f_x(\Psi)}q^{R(\Psi)+|\rho_\Psi(\mm)|};q^2)_\infty}\,,
	\end{split}\label{SCI:0-1}
\end{equation}
where $r_G$ denotes the rank of the gauge group and the mixed CS levels are turned on in general. The underlined extra phase terms due to the replacement $(-1)^F\to(-1)^{2j_3}$ in the trace formula (\ref{SCI:tr}) can be written as
\begin{equation}
	\zeta=\prod_{\ell,n=1}^{r_G}(-1)^{k_{\ell n}\mm_{\ell}\mm_{n}}\times\prod_{\Psi}\prod_{\rho_\Psi}(-1)^{-\fft12\rho_\Psi(\mm)|\rho_\Psi(\mm)|}\,.\label{zeta:1}
\end{equation}
Introducing the explicit linear form of the weights, with $\mathfrak{h}_G$ the Cartan subalgebra of the gauge group $G$,
\begin{equation}
	\rho_\Psi(X)=\sum_{\ell=1}^{r_G}(\rho_\Psi)_{\ell}X_{\ell}\,, \qquad(X\in\mathfrak{h}_G\,,~(\rho_\Psi)_{\ell}\in\mathbb{Z})\,,
\end{equation}
and using that the gauge magnetic fluxes are integer quantized, $\mm_{\ell}\in\mathbb{Z}$, one can rewrite the phase factor (\ref{zeta:1}) as
\begin{equation}
\begin{split}
	\zeta&=\prod_{\ell,n=1}^{r_G}(-1)^{k_{\ell n}\mm_{\ell}\mm_{n}}\times\prod_{\Psi}\prod_{\rho_\Psi}(-1)^{\fft12\rho_\Psi(\mm)^2}(-1)^{\fft12(\rho_\Psi(\mm)+|\rho_\Psi(\mm)|)}\\
	&=\prod_{\ell,n=1}^{r_G}(-1)^{\big(k_{\ell n}+\fft12\sum_\Psi\sum_{\rho_\Psi}(\rho_\Psi)_\ell(\rho_\Psi)_n\big)\mm_{\ell}\mm_{n}}\times\prod_{\Psi}\prod_{\rho_\Psi}(-1)^{\fft12(\rho_\Psi(\mm)+|\rho_\Psi(\mm)|)}\,.
\end{split}\label{zeta:2}
\end{equation}
The shifted CS levels are integer quantized as \cite{Aharony:1997bx}\footnote{In \cite{Closset:2017zgf,Closset:2019hyt} the authors take the U(1)$_{-\fft12}$ quantization for chiral multiplets so that the bare CS levels are integer quantized. In that case the localization formula for the 1-loop contribution of a chiral multiplet to the SCI also slightly changes \cite{Closset:2019hyt} compared to the convention of \cite{Aharony:2013dha,Aharony:2013kma,Garozzo:2019ejm}. In the latter convention we followed in this paper, the integer quantized CS levels correspond to the shifted ones (\ref{k:shifted}).}
\begin{equation}
	k_{\ell n}+\fft12\sum_\Psi\sum_{\rho_\Psi}(\rho_\Psi)_\ell(\rho_\Psi)_n\in\mathbb{Z}\,,\label{k:shifted}
\end{equation}
and therefore the phase factor (\ref{zeta:2}) can be simplified further as 
\begin{equation}
	\zeta=\prod_{\ell=1}^{r_G}(-1)^{\big(k_{\ell\ell}+\fft12\sum_\Psi\sum_{\rho_\Psi}(\rho_\Psi)_\ell(\rho_\Psi)_\ell\big)\mm_{\ell}}\times\prod_{\Psi}\prod_{\rho_\Psi}(-1)^{\fft12(\rho_\Psi(\mm)+|\rho_\Psi(\mm)|)}\,.\label{zeta:3}
\end{equation}

Substituting the rewritten phase factor (\ref{zeta:3}) back into the localization formula (\ref{SCI:0-1}) and specializing it to the class of $\mN=2$ SCFTs of our interest described in Section \ref{sec:general:summary} where the subscript $\ell$ labeling the Cartan generators of the gauge group is replaced with the pair $(r,i)$, we obtain
\begin{equation}
	\begin{split}
		\mI(q,\boldsymbol{\xi})&=\fft{1}{(N!)^p}\sum_{\mm_1,\cdots,\mm_p\in\mathbb Z^N}\oint\Bigg[\prod_{r=1}^p\prod_{i=1}^N\fft{dz_{r,i}}{2\pi \ri z_{r,i}}z_{r,i}^{k_r\mm_{r,i}}(-1)^{\big(k_r+\fft12\sum_\Psi\sum_{\rho_\Psi}(\rho_\Psi)_{r,i}(\rho_\Psi)_{r,i}\big)\mm_{r,i}}\xi_{T_r}^{\mm_{r,i}}\Bigg]\\
		&\quad\times\prod_{r=1}^p\prod_{i\neq j}^Nq^{-\fft12|\mm_{r,i}-\mm_{r,j}|}\left(1-z_{r,i}z_{r,j}^{-1}q^{|\mm_{r,i}-\mm_{r,j}|}\right)\\
		&\quad\times\prod_{\Psi}\prod_{\rho_\Psi}(-1)^{\fft12(\rho_\Psi(\mm)+|\rho_\Psi(\mm)|)}\left(q^{1-R(\Psi)}e^{-\ri\rho_\Psi(h)}\prod_{x}\xi_x^{-f_x(\Psi)}\right)^{\fft12|\rho_\Psi(\mm)|}\\
		&\kern5em\times\fft{(e^{-\ri\rho_\Psi(h)}\prod_{x}\xi_x^{-f_x(\Psi)}q^{2-R(\Psi)+|\rho_\Psi(\mm)|};q^2)_\infty}{(e^{\ri\rho_\Psi(h)}\prod_{x}\xi_x^{f_x(\Psi)}q^{R(\Psi)+|\rho_\Psi(\mm)|};q^2)_\infty}\,.
	\end{split}\label{SCI:0-2}
\end{equation}
Note that we have turned off mixed CS levels between gauge/global symmetries except the unit mixed CS levels between gauge and U(1) topological symmetries. Finally, we redefine the fugacities associated with topological symmetries as 
\begin{equation}
	\xi_{T_r}\quad\to\quad\xi_{T_r}(-1)^{-\big(k_r+\fft12\sum_\Psi\sum_{\rho_\Psi}(\rho_\Psi)_{r,i}(\rho_\Psi)_{r,i}\big)}\,,\label{xiTr:redefine}
\end{equation}
upon which (\ref{SCI:0-2}) yields the localization formula (\ref{SCI:1}) in the main text. Note that the redefinition (\ref{xiTr:redefine}) is allowed if the shifted CS level $k_r+\fft12\sum_\Psi\sum_{\rho_\Psi}(\rho_\Psi)_{r,i}(\rho_\Psi)_{r,i}$ is independent of the subscript $i\in\{1,\cdots,N\}$, which is indeed the case for $\mN=2$ SCFTs involving chiral multiplets in the following representations:
\begin{itemize}
	\item (Anti)-fundamental representation $\Psi_s$ with $N$ weights $(\rho_{\Psi_s}^{(j)})_{r,i}=\pm\delta_{s,r}\delta_{j,i}$ where $j\in\{1,\cdots,N\}$,
	
	\item Adjoint representation $\Psi_{(s,s)}$ with $N^2$ weights $(\rho_{\Psi_{(s,s)}}^{(j,k)})_{r,i}=\delta_{s,r}(\delta_{j,i}-\delta_{k,i})$ where $j,k\in\{1,\cdots,N\}$,
	
	\item Bi-fundamental representation $\Psi_{(s,t)}$ $(s\neq t)$ with $N^2$ weights $(\rho_{\Psi_{(s,t)}}^{(j,k)})_{r,i}=\delta_{s,r}\delta_{j,i}-\delta_{t,r}\delta_{k,i}$ where $j,k\in\{1,\cdots,N\}$.
\end{itemize}
%

%%%%%
\section{Cardy-like expansion of the SCI }\label{app:Cardy}
%%%%%
In this appendix we provide details in the derivation of the Cardy-like limit of the SCI (\ref{SCI:Cardy:3}). To begin with, one can expand the SCI~(\ref{SCI:3}) in the Cardy-like limit~(\ref{Cardy-like}) using the asymptotic expansion (\ref{poch:asymp}) as \cite{Choi:2019dfu,Choi:2019zpz}
\begin{equation}
	\begin{split}
		\mI(\omega,\Delta,\mn)&=\fft{1}{(N!)^p}(-1)^{\fft{pN(N-1)}{2}}\sum_{\mm_1,\cdots,\mm_p\in\mathbb Z^N}\oint_{|s_{r,i}|=e^{-\ri\pi\omega\mm_{r,i}}}\bigg(\prod_{r=1}^p\prod_{i=1}^N\fft{ds_{r,i}}{2\pi \ri s_{r,i}}\bigg)\\
		&\quad\times\exp[-\fft{\ri}{2\pi\omega}\mW[U;\Delta,\mn,\omega,\ell]+\fft{\ri}{2\pi\omega}\mW[-\bU;-\Delta,\mn,-\omega,-\ell]+\mO(\omega)]\,,
	\end{split}\label{SCI:Cardy:1}
\end{equation}
where we have introduced a holomorphic effective potential
\begin{equation}
	\begin{split}
		&\mW[U;\Delta,\mn,\omega,\ell]\\
		&=\sum_{r=1}^p\sum_{i=1}^N\bigg[\fft12k_rU_{r,i}^2-\pi(2\ell_{r,i}-\Delta_{T_r}+\omega\mft_r)U_{r,i}\bigg]-\ri\pi\omega\sum_{r=1}^p\sum_{i\neq j}^N\text{Li}_1(e^{\ri(U_{r,j}-U_{r,i})})\\
		&+\sum_\Psi\sum_{\rho_\Psi}\bigg[\fft14\rho_\Psi(U)^2+\fft{\pi\Delta_\Psi}{2}\rho_\Psi(U)-\fft{1}{2}\pi\omega(1-\mn_\Psi)\big(\rho_\Psi(U)+\pi\Delta_\Psi\big)-\text{Li}_2(e^{\ri\rho_\Psi(U)}y_\Psi q^{-1+\mn_\Psi})\bigg]\,.
	\end{split}\label{mW}
\end{equation}
In the last line of (\ref{mW}) we have included an extra term proportional to $\sim\omega(1-\mn_\Psi) \Delta_\Psi$ which cancels in the full exponent of (\ref{SCI:Cardy:1}) but is useful later. Note that the holomorphic effective potential (\ref{mW}) is expanded in the Cardy-like limit as in (\ref{mW:Cardy}) using the asymptotic expansion (\ref{poch:asymp}). Now using the complex conjugate relations
\begin{equation}
	\begin{split}
		\overline{\mW^{(0)}[U;\Delta,\ell]}&=\mW^{(0)}[-\bU;-\Delta,-\ell]\,,\\
		\overline{\mW^{(1)}[U;\Delta,\mn]}&=\mW^{(1)}[-\bU;-\Delta,\mn]\,,
	\end{split}\label{cc}
\end{equation}
for the expanded effective potential (\ref{mW:Cardy}), one can rewrite (\ref{SCI:Cardy:1}) as
\begin{equation}
	\begin{split}
		\mI(\omega,\Delta,\mn)&=\fft{1}{(N!)^p}(-1)^{\fft{pN(N-1)}{2}}\sum_{\mm_1,\cdots,\mm_p\in\mathbb Z^N}\oint_{|s_{r,i}|=e^{-\ri\pi\omega\mm_{r,i}}}\bigg(\prod_{r=1}^p\prod_{i=1}^N\fft{ds_{r,i}}{2\pi \ri s_{r,i}}\bigg)\\
		&\quad\times\exp[\fft{1}{\pi\omega}\Im\mW^{(0)}[U;\Delta,\ell]+2\Re\mW^{(1)}[U;\Delta,\mn]+\mO(\omega)]\,.
	\end{split}\label{SCI:Cardy:2}
\end{equation}
To simplify (\ref{SCI:Cardy:2}) further in the Cardy-like limit, we replace the sums over gauge magnetic fluxes with the integrals as
\begin{equation}
	\begin{split}
		&\sum_{\mm_1,\cdots,\mm_p\in\mathbb Z^N}\oint_{|s_{r,i}|=e^{-\ri\pi\omega\mm_{r,i}}}\bigg(\prod_{r=1}^p\prod_{i=1}^N\fft{ds_{r,i}}{2\pi \ri s_{r,i}}\bigg)(\cdots)\\
		&=\int_{\mathbb C^{pN}}\bigg(\prod_{r=1}^p\prod_{i=1}^N\fft{dU_{r,i}d\bU_{r,i}}{-4\ri\pi^2\omega}\bigg)(\cdots)\left(1+\mathcal O(\omega)\right)\,,\label{sum:to:integral}
	\end{split}
\end{equation}
following \cite{Choi:2019dfu,Choi:2019zpz} and Appendix C.1 of \cite{Bobev:2022wem} based on the Euler-Maclaurin formula. Applying the replacement (\ref{sum:to:integral}) to (\ref{SCI:Cardy:2}) we arrive at (\ref{SCI:Cardy:3}).

%%%%%
\section{Revisiting ABJM/ADHM theories }\label{app:ABJMADHM}
%%%%%
In this Appendix we revisit the Cardy-like expansion of the SCI for ABJM/ADHM theories studied in \cite{Bobev:2022wem} following the generic $\mN=2$ conventions spelled out in Section~\ref{sec:general}. We explicitly show that some technical differences in the intermediate steps of the calculation do not affect the final results of \cite{Bobev:2022wem}.

%%%%%
\subsection{ABJM theory}\label{app:ABJMADHM:ABJM}
%%%%%
The generic Bethe potential (\ref{V}) and the generic BA formula for the TTI (\ref{TTI:3}) for the ABJM theory can be read off from the ones for the $N^{0,1,0}$ theory, (\ref{V:N010}) and (\ref{TTI:N010}), simply by setting $r_1=r_2=0$. One can check that the resulting expressions match the Bethe potential\footnote{To match the Bethe potentials up to gauge holonomy independent terms, we have used the inversion formula (\ref{polylog:inversion}) under the assumption
\begin{equation}
	0<\Re[u_{i}-\tu_{j}+\pi\Delta_{1,2}]<2\pi\,.
\end{equation}
} and the TTI of the ABJM theory in \cite{Bobev:2022eus} respectively after the identifications
\begin{equation}
	\begin{split}
		(u_{1,i},u_{2,i})&=(u_i,\tu_i)\,,\\
		(\Delta_{m,1},\Delta_{m,2})&=(0,0)\,,\\
		(2n_{1,i},2n_{2,i})&=(2n_i-\fft{1-(-1)^N}{2}+N,2\tn_i-\fft{1-(-1)^N}{2}+N)\,,\\
		(\mft_1,\mft_2)&=(0,0)\,,
	\end{split}
\end{equation}\label{ABJM:identification}
and the constraints
\begin{equation}
	\sum_{a=1}^4\Delta_a=2\,,\qquad \qquad\sum_{a=1}^4\mn_a=2\,.\label{ABJM:constraints}
\end{equation}
Therefore we can use the numerical BAE solutions $\{u_\star\}$ constructed in \cite{Bobev:2022eus} for the choice of integers
\begin{equation}
	(n_i,\tn_j)=(1-i,j-N)\,,\label{ABJM:integer}
\end{equation}
for various configurations satisfying the constraints (\ref{ABJM:constraints}). Furthermore we can use the closed form expression for the ABJM TTI derived by substituting those numerical BAE solutions to the BA formula which reads \cite{Bobev:2022eus}
\begin{equation}
	\begin{split}
		\log Z_\text{ABJM}(\Delta,\mn)&=-\fft{\pi\sqrt{2k\Delta_1\Delta_2\Delta_3\Delta_4}}{3}\sum_{a=1}^4\fft{\mn_a}{\Delta_a}\left(\hat N_\Delta^\fft32-\fft{\mathfrak c_a}{k}\hat N_\Delta^\fft12\right)\\
		&\quad-\fft12\log\hat N_\Delta+\hat f_0(k,\Delta,\mn)+\hat{f}_\text{np}(N,k,\Delta,\mn) \, ,
	\end{split}\label{TTI:ABJM:result}
\end{equation}
where we have defined
\begin{subequations}
	\begin{align}
		\hat N_\Delta&=N-\fft{k}{24}+\fft{1}{12k}\sum_{a=1}^4\fft{1}{\Delta_a} \, , \\
		\mathfrak c_a&=\fft{\prod_{b\neq a}(\Delta_a+\Delta_b)}{8\Delta_1\Delta_2\Delta_3\Delta_4}\sum_{b\neq a}\Delta_b \, .
	\end{align}\label{ABJM:details}%
\end{subequations}

Now we move on to the SCI of the ABJM theory. As in the various examples studied in Section~\ref{sec:ex}, it suffices to determine the closed form expression for the Bethe potential using various BAE solutions $\{u_\star\}$ in \cite{Bobev:2022eus}. We obtained
\begin{equation}
	\Im\mV_\text{ABJM}[u_\star;\Delta,n]=2\pi\left[\fft{\pi\sqrt{2k\Delta_1\Delta_2\Delta_3\Delta_4}}{3}\hat N_{k,\Delta}^\fft32+\hat g_0(k,\Delta)+\hat{g}_\text{np}(N,k,\Delta)\right]\,,\label{mW:Cardy:0:ABJM:saddle} 
\end{equation}
which is exactly the same as the result of \cite{Bobev:2022wem}. In other words, we have shown that the gauge holonomy independent difference between the ABJM Bethe potential (\ref{V:N010}) with $r_1=r_2=0$ and the expression used in \cite{Bobev:2022eus} becomes real at the BAE solution and therefore does not affect the result (\ref{mW:Cardy:0:ABJM:saddle}). 

Substituting the TTI (\ref{TTI:ABJM:result}) and the Bethe potential (\ref{mW:Cardy:0:ABJM:saddle}) into the Cardy-like expansion (\ref{summary:improved}), we arrive at the result of \cite{Bobev:2022wem}, which we repeat here for completeness
\begin{equation}
	\begin{split}
		\log\mI_\text{ABJM}(\omega,\Delta,\mn)&=-\fft{2}{\omega}\left[\fft{\pi\sqrt{2k\Delta_1\Delta_2\Delta_3\Delta_4}}{3}\hat N_{k,\Delta}^\fft32+\hat g_0(k,\Delta)+\hat{g}_\text{np}(N,k,\Delta)\right]\\
		&\quad+\Bigg[-\fft{\pi\sqrt{2k\Delta_1\Delta_2\Delta_3\Delta_4}}{3}\sum_{a=1}^4\fft{\mn_a}{\Delta_a}\left(\hat N_{k,\Delta}^\fft32-\fft{\mathfrak{c}_a(\Delta)}{k}\hat N_{k,\Delta}^\fft12\right)\\
		&\kern3em-\fft12\log\hat N_{k,\Delta}+\hat f_0(k,\Delta,\mn)+\hat{f}_\text{np}(N,k,\Delta,\mn)\Bigg]+\mathcal O(\omega)\,.
	\end{split}\label{SCI:ABJM}
\end{equation}
These calculations confirm that the approach based on the generic $\mN=2$ conventions is consistent with the previous analysis for the ABJM theory in \cite{Bobev:2022wem}.

%%%%%
\subsection{ADHM theory}\label{app:ABJMADHM:ADHM}
%%%%%
The generic Bethe potential (\ref{V}) and the BA formula for the TTI (\ref{TTI:3}) for the ADHM theory are equivalent to those for the $V^{5,2}$ theory given in Section~\ref{sec:ex:V52} and match the ones in \cite{Bobev:2023lkx} under the same identifications (\ref{V52:identification}) but different constraints
\begin{equation}
	\begin{split}
		\sum_{I=1}^3\Delta_I&=\Delta_3+\Delta_q+\Delta_{\tq}=2\,,\\
		\sum_{I=1}^3\mn_I&=\mn_3+\mn_q+\mn_{\tq}=2\,.
	\end{split}\label{ADHM:constraints}
\end{equation}
Hence we can use the numerical BAE solutions $\{u_\star\}$ constructed in \cite{Bobev:2023lkx} with the choice of integers $n_i=\lfloor\fft{N+1}{2}\rfloor-i$ for various configurations satisfying the constraints (\ref{ADHM:constraints}). Furthermore we can use the closed form expression for the ADHM TTI derived in \cite{Bobev:2023lkx} 
\begin{equation}
	\begin{split}
		&\Re\log Z_\text{ADHM}(\Delta,\mn)\\
		&=-\fft{\pi\sqrt{2N_f\tilde\Delta_1\tilde\Delta_2\tilde\Delta_3\tilde\Delta_4}}{3}\sum_{a=1}^4\tilde\mn_a\left[\fft{1}{\tilde\Delta_a}(\hat N_{N_f,\tilde\Delta})^\fft32+\bigg(\mathfrak c_a(\tilde\Delta)N_f+\fft{\mathfrak d_a(\tilde\Delta)}{N_f}\bigg)(\hat N_{N_f,\tilde\Delta})^\fft12\right]\\
		&\quad-\fft12\log\hat N_{N_f,\tilde\Delta}+\hat f_0(N_f,\tilde\Delta,\tilde\mn)+\hat{f}_\text{np}(N,N_f,\tDelta,\tmn) \, ,
	\end{split}\label{TTI:ADHM:result}
\end{equation}
where the shifted $N$ parameter is given by (\ref{V52:details}) and we have also defined
\begin{subequations}
	\begin{align}
		\mathfrak c_a(\tilde\Delta)&=\bigg(\!\!-\fft{1}{\tilde\Delta_1}\fft{(\tilde\Delta_2+\tilde\Delta_3+\tilde\Delta_4)(\tilde\Delta_1+\tilde\Delta_2)}{8\tilde\Delta_1\tilde\Delta_2},-\fft{1}{\tilde\Delta_2}\fft{(\tilde\Delta_1+\tilde\Delta_3+\tilde\Delta_4)(\tilde\Delta_1+\tilde\Delta_2)}{8\tilde\Delta_1\tilde\Delta_2},\nn\\
		&\quad~-\fft{\tilde\Delta_3+\tilde\Delta_4}{8\tilde\Delta_1\tilde\Delta_2},-\fft{\tilde\Delta_3+\tilde\Delta_4}{8\tilde\Delta_1\tilde\Delta_2}\bigg)\,,\\
		\mathfrak d_a(\tilde\Delta)&=\bigg(\!\!-\fft{(\tilde\Delta_1+\tilde\Delta_2)(\tilde\Delta_2+\tilde\Delta_3+\tilde\Delta_4)(\tilde\Delta_1+\tilde\Delta_3+\tilde\Delta_4)}{8\tilde\Delta_1\tilde\Delta_2\tilde\Delta_3\tilde\Delta_4},\nn\\
		&\quad~-\fft{(\tilde\Delta_1+\tilde\Delta_2)(\tilde\Delta_2+\tilde\Delta_3+\tilde\Delta_4)(\tilde\Delta_1+\tilde\Delta_3+\tilde\Delta_4)}{8\tilde\Delta_1\tilde\Delta_2\tilde\Delta_3\tilde\Delta_4},\nn\\
		&\quad~-\fft{1}{\tilde\Delta_3}\fft{(\tilde\Delta_3+\tilde\Delta_4)((\tDelta_1+\tDelta_2)(\tDelta_2+\tDelta_3)(\tDelta_3+\tDelta_1)+(\tDelta_1\tDelta_2+\tDelta_2\tDelta_3+\tDelta_3\tDelta_1)\tDelta_4)}{8\tilde\Delta_1\tilde\Delta_2\tilde\Delta_3\tilde\Delta_4},\nn\\
		&\quad~-\fft{1}{\tilde\Delta_4}\fft{(\tilde\Delta_3+\tilde\Delta_4)((\tDelta_1+\tDelta_2)(\tDelta_2+\tDelta_4)(\tDelta_4+\tDelta_1)+(\tDelta_1\tDelta_2+\tDelta_2\tDelta_4+\tDelta_4\tDelta_1)\tDelta_3)}{8\tilde\Delta_1\tilde\Delta_2\tilde\Delta_3\tilde\Delta_4}\bigg)\,,
	\end{align}\label{ADHM:details}%
\end{subequations}
in addition to (\ref{V52:details:2}).

Now we move on to the SCI of the ADHM theory. As in Section~\ref{sec:ex}, it suffices to determine the closed form expression for the Bethe potential using various BAE solutions $\{u_\star\}$ in \cite{Bobev:2023lkx}. We obtained
\begin{equation}
	\Im\mV_\text{ADHM}[u_\star;\Delta,n]=2\pi\left[\fft{\pi\sqrt{2N_f\tDelta_1\tDelta_2\tDelta_3\tDelta_4}}{3}\hat N_{N_f,\Delta}^\fft32+\hat g_0(N_f,\Delta)+\hat{g}_\text{np}(N,N_f,\tDelta)\right]\,,\label{mW:Cardy:0:ADHM:saddle} 
\end{equation}
which is exactly the same as the result of \cite{Bobev:2022wem}. As in the ABJM case discussed above, this implies that the gauge holonomy independent difference between the Bethe potential (\ref{V:V52}) and the one used in \cite{Bobev:2023lkx} becomes real at the BAE solution and therefore does not affect the result (\ref{mW:Cardy:0:ADHM:saddle}). 

Substituting the TTI (\ref{TTI:ADHM:result}) and the Bethe potential (\ref{mW:Cardy:0:ADHM:saddle}) into the Cardy-like expansion (\ref{summary:improved}), we arrive at the result of \cite{Bobev:2022wem}, which we repeat here for completeness
\begin{equation}
	\begin{split}
		&\log\mI_\text{ADHM}(\omega,\Delta,\mn)\\
		&=-\fft{2}{\omega}\left[\fft{\pi\sqrt{2N_f\tDelta_1\tDelta_2\tDelta_3\tDelta_4}}{3}\hat N_{N_f,\Delta}^\fft32+\hat g_0(N_f,\Delta)+\hat{g}_\text{np}(N,N_f,\tDelta)\right]\\
		&\quad+\Bigg[-\fft{\pi\sqrt{2N_f\tDelta_1\tDelta_2\tDelta_3\tDelta_4}}{3}\sum_{a=1}^4\tmn_a\left[\fft{1}{\tDelta_a}\hat N_{N_f,\tDelta}^\fft32+\bigg(\mathfrak{c}_a(\tDelta)N_f+\fft{\mathfrak{d}_a(\tDelta)}{N_f}\bigg)\hat N_{N_f,\tDelta}^\fft12\right]\\
		&\kern3em-\fft12\log\hat N_{N_f,\tDelta}+\hat f_0(N_f,\tDelta,\tmn)+\hat{f}_\text{np}(N,N_f,\tDelta,\tmn)\Bigg]+\ri\varphi+\mathcal O(\omega)\,.
	\end{split}
\end{equation}
This again confirms that the approach based on the generic $\mN=2$ conventions is consistent with the previous analysis for the ADHM theory in \cite{Bobev:2022wem}.

%%%%%
\section{Conventions and Bethe potential }\label{app:Bethe}
%%%%%
In this Appendix we first provide a relation between the generic $\mN=2$ TTI conventions introduced in Section~\ref{sec:general:TTI} and the conventions of \cite{Bobev:2023lkx} for various $\mN=2$ holographic SCFTs studied in Section~\ref{sec:ex}, which allows us to employ the results of \cite{Bobev:2023lkx} in analyzing the Cardy-like expansion of the SCI. We then present numerical data that supports the analytic expressions for the Bethe potentials associated with the TTI.

%%%%%
\subsection{$N^{0,1,0}$ theory}\label{app:Bethe:N010}
%%%%%
The Bethe potential (\ref{V:N010}) and the TTI (\ref{TTI:N010}) of the $N^{0,1,0}$ theory can be identified with the expressions presented in \cite{Bobev:2023lkx} after the identifications
\begin{equation}
	\begin{split}
		(u_{1,i},u_{2,i})&=(u_i,\tu_i)\,,\\
		(\Delta_{m,1},\Delta_{m,2})&=(N+1,-N-1)\,,\\
		(n_{1,i},n_{2,i})&=(n_i+N,\tn_i+N)\,,\\
		(\mft_1,\mft_2)&=(0,0)\,.
	\end{split}\label{N010:identification}
\end{equation}
To match the Bethe potentials, we have used the inversion formula (\ref{polylog:inversion}) with the assumptions\footnote{Note that the Bethe potential is not written explicitly for the $N^{0,1,0}$ TTI in \cite{Bobev:2023lkx} but it can be deduced easily from the BAE modulo gauge holonomy independent terms.}
\begin{equation}
	0<\Re[u_{i}-\tu_{j}+\pi\Delta_{1,2}]<2\pi\quad\&\quad 0<\Re[u_{i}+\pi\Delta_{q_1}],\Re[\tu_{i}+\pi\Delta_{q_2}]<2\pi\,.
\end{equation}
In the comparison of the TTI, we observed a slight phase difference between (\ref{TTI:N010}) and the TTI expression in \cite{Bobev:2023lkx}, which originates from a subtle branch choice for flavor fugacities. It does not affect the real part of the logarithm of the TTI that is relevant in the index relation (\ref{summary:improved}), however, and therefore the Cardy-like expansion of the $N^{0,1,0}$ SCI will not be affected by this minor phase difference. Hence we ignore this phase difference throughout the paper and similarly in other holographic SCFTs. 

\medskip

We now turn to the numerical data that supports the all-order $1/N$ expansion of the $N^{0,1,0}$ Bethe potential given in (\ref{mW:Cardy:0:N010:saddle}). The list of $(k,r)$-configurations for which we confirmed the analytic expression in (\ref{mW:Cardy:0:N010:saddle}) is given as follows: 
\begin{equation}
	\begin{split}
		k\in\{1,2,3,4\}\,,\qquad \fft{r}{k}\in\left\{\fft12,\fft23,1,\fft32,2,3\right\}\,.
	\end{split}\label{N010:case}
\end{equation}
For the above list of $(k,r)$-configurations, we estimate the numerical coefficients
\begin{equation}
	\hat g_{3/2}^\text{(lmf)}(k,r)\qquad\&\qquad\hat g_0^\text{(lmf)}(k,r)
\end{equation}
together with the associated standard errors $\sigma_{3/2}$ and $\sigma_0$ in the \texttt{LinearModelFit} for the $N^{0,1,0}$ Bethe potential. Namely, we evaluate
\begin{equation}
	\fft{1}{2\pi}\Im\mV_{N^{0,1,0}}[u_\star;\Delta,n]\Big|_\text{(\ref{N010:special})}=\hat g_{3/2}^\text{(lmf)}(k,r)\hat N_{k,r}^\fft32+\hat g_0^\text{(lmf)}(k,r)\,,\label{N010:lmf}
\end{equation}
following the numerical BAE solutions constructed in \cite{Bobev:2023lkx} for $N=101\sim301~(\text{in steps of }10)$ at \texttt{WorkingPrecision} $200$. The leading order coefficient is then compared with the corresponding analytic expression in (\ref{mW:Cardy:0:N010:saddle}), namely
\begin{equation}
	\begin{split}
		\hat g_{3/2}(k,r)=\fft{\pi(k+r)}{6\sqrt{2k+r}}\,,
	\end{split}
\end{equation}
by calculating the error ratio
\begin{equation}
	R_{3/2}(k,r) = \fft{\hat g_{3/2}^\text{(lmf)}(k,r)-\hat g_{3/2}(k,r)}{\hat g_{3/2}(k,r)}\,.
\end{equation}
The following tables summarize the numerical data described above.

\medskip

%%%%%
\noindent$\boldsymbol{k=r}$
%%%%%
%
\begin{center}
	\footnotesize
	\begin{tabular}{ |c||c|c|c|c| } 
		\hline
		& $R_{3/2}$ & $\sigma_{3/2}$ & $\hat g_0^\text{(lmf)}$ & $\sigma_0$\\
		\hline\hline
		$k=1$ & $5.725{\times}10^{-20}$ & $1.802{\times}10^{-20}$ & $-0.16014196980357514241$ & $5.808{\times}10^{-17}$ \\
		\hline
		$k=2$ & $-1.578{\times}10^{-26}$ & $7.839{\times}10^{-27}$ & $-0.21428948039679424932$ & $2.526{\times}10^{-23}$\\
		\hline
		$k=3$ & $5.840{\times}10^{-15}$ & $2.500{\times}10^{-15}$ & $-0.32655189751806194061$ & $8.058{\times}10^{-12}$\\ 
		\hline
		$k=4$ & $1.736{\times}10^{-13}$ & $7.911{\times}10^{-14}$ & $-0.48717531374612447878$ & $2.551{\times}10^{-10}$\\ 
		\hline
	\end{tabular}
\end{center}

\medskip

%%%%%
\noindent$\boldsymbol{2k=r}$
%%%%%
%
\begin{center}
	\footnotesize
	\begin{tabular}{ |c||c|c|c|c| } 
		\hline
		& $R_{3/2}$ & $\sigma_{3/2}$ & $\hat g_0^\text{(lmf)}$ & $\sigma_0$\\
		\hline\hline
		$k=1$ & $1.410{\times}10^{-14}$ & $4.528{\times}10^{-15}$ & $-0.22242988776920419681$ & $1.460{\times}10^{-11}$ \\
		\hline
		$k=2$ & $4.378{\times}10^{-12}$ & $1.620{\times}10^{-12}$ & $-0.50364353625504316942$ & $5.226{\times}10^{-9}$\\
		\hline
		$k=3$ & $1.216{\times}10^{-10}$ & $4.816{\times}10^{-11}$ & $-0.98356557594764269955$ & $1.556{\times}10^{-7}$\\ 
		\hline
		$k=4$ & $9.830{\times}10^{-10}$ & $4.063{\times}10^{-10}$ & $-1.6571101147599629963$ & $1.315{\times}10^{-6}$\\ 
		\hline
	\end{tabular}
\end{center}

\medskip

%%%%%
\noindent$\boldsymbol{k=2r}$
%%%%%
%
\begin{center}
	\footnotesize
	\begin{tabular}{ |c||c|c|c|c| } 
		\hline
		& $R_{3/2}$ & $\sigma_{3/2}$ & $\hat g_0^\text{(lmf)}$ & $\sigma_0$\\
		\hline\hline
		$k=1$ & $2.009{\times}10^{-18}$ & $4.850{\times}10^{-19}$ & $-0.15676419358385312020$ & $1.564{\times}10^{-15}$ \\
		\hline
		$k=2$ & $-1.037{\times}10^{-21}$ & $3.873{\times}10^{-22}$ & $-0.15305814860409592117$ & $1.247{\times}10^{-18}$\\
		\hline
		$k=3$ & $5.569{\times}10^{-20}$ & $2.355{\times}10^{-20}$ & $-0.18114828105867926447$ & $7.583{\times}10^{-17}$\\ 
		\hline
		$k=4$ & $1.525{\times}10^{-17}$ & $6.984{\times}10^{-18}$ & $-0.22615656081972779158$ & $2.248{\times}10^{-14}$\\ 
		\hline
	\end{tabular}
\end{center}

\medskip

%%%%%
\noindent$\boldsymbol{3k=2r}$
%%%%%
%
\begin{center}
	\footnotesize
	\begin{tabular}{ |c||c|c|c|c| } 
		\hline
		& $R_{3/2}$ & $\sigma_{3/2}$ & $\hat g_0^\text{(lmf)}$ & $\sigma_0$\\
		\hline\hline
		$k=1$ & $5.496{\times}10^{-17}$ & $1.783{\times}10^{-17}$ & $-0.18315250021650044933$ & $5.747{\times}10^{-14}$ \\
		\hline
		$k=2$ & $3.819{\times}10^{-14}$ & $1.463{\times}10^{-14}$ & $-0.33156998734746390156$ & $4.717{\times}10^{-11}$\\
		\hline
		$k=3$ & $2.571{\times}10^{-12}$ & $1.068{\times}10^{-12}$ & $-0.59423904669712970624$ & $3.448{\times}10^{-9}$\\ 
		\hline
		$k=4$ & $3.351{\times}10^{-11}$ & $1.466{\times}10^{-11}$ & $-0.96428722814277342203$ & $4.735{\times}10^{-8}$\\ 
		\hline
	\end{tabular}
\end{center}

\medskip

%%%%%
\noindent$\boldsymbol{2k=3r}$
%%%%%
%
\begin{center}
	\footnotesize
	\begin{tabular}{ |c||c|c|c|c| } 
		\hline
		& $R_{3/2}$ & $\sigma_{3/2}$ & $\hat g_0^\text{(lmf)}$ & $\sigma_0$\\
		\hline\hline
		$k=1$ & $3.269{\times}10^{-19}$ & $8.755{\times}10^{-20}$ & $-0.15532082590294874648$ & $2.822{\times}10^{-16}$ \\
		\hline
		$k=2$ & $-6.275{\times}10^{-20}$ & $2.394{\times}10^{-20}$ & $-0.16709079161756907785$ & $7.712{\times}10^{-17}$\\
		\hline
		$k=3$ & $7.998{\times}10^{-18}$ & $3.421{\times}10^{-18}$ & $-0.21595042199749374794$ & $1.102{\times}10^{-14}$\\ 
		\hline
		$k=4$ & $7.039{\times}10^{-16}$ & $3.239{\times}10^{-16}$ & $-0.28910459214092956841$ & $1.043{\times}10^{-12}$\\ 
		\hline
	\end{tabular}
\end{center}

\medskip

%%%%%
\noindent$\boldsymbol{3k=r}$
%%%%%
%
\begin{center}
	\footnotesize
	\begin{tabular}{ |c||c|c|c|c| } 
		\hline
		& $R_{3/2}$ & $\sigma_{3/2}$ & $\hat g_0^\text{(lmf)}$ & $\sigma_0$\\
		\hline\hline
		$k=1$ & $1.273{\times}10^{-11}$ & $3.840{\times}10^{-12}$ & $-0.34513332066468836843$ & $1.239{\times}10^{-8}$ \\
		\hline
		$k=2$ & $1.081{\times}10^{-9}$ & $3.631{\times}10^{-10}$ & $-1.0105562008492962901$ & $1.174{\times}10^{-6}$\\
		\hline
		$k=3$ & $1.215{\times}10^{-8}$ & $4.297{\times}10^{-9}$ & $-2.1264119370284434162$ & $1.392{\times}10^{-5}$\\ 
		\hline
		$k=4$ & $5.688{\times}10^{-8}$ & $2.079{\times}10^{-8}$ & $-3.6898237530453026380$ & $6.753{\times}10^{-5}$\\ 
		\hline
	\end{tabular}
\end{center}

\medskip

To argue that the \texttt{LinearModelFit} in (\ref{N010:lmf}) captures the perturbative expansion of the Bethe potential completely up to non-perturbative corrections of order $\mO(e^{-\sqrt{N}})$, one can follow the numerical technique introduced in \cite{Bobev:2022eus} for the non-perturbative analysis. Based on that approach, we confirmed that (\ref{N010:lmf}) is indeed exact up to exponentially suppressed non-perturbative corrections and did a similar check for the $V^{5,2}$ and $Q^{1,1,1}$ theories. In the discussion below  we omit the details of the analysis of the non-perturbative corrections for these two other models since it is parallel to the analysis in \cite{Bobev:2022eus}.

%%%%%
\subsection{$V^{5,2}$ theory}\label{app:Bethe:V52}
%%%%%
The Bethe potential (\ref{V:V52}) and the TTI (\ref{TTI:V52}) of the $V^{5,2}$ theory can be identified with the expressions presented in \cite{Bobev:2023lkx} after the substitutions
\begin{equation}
	\begin{split}
		u_{1,i}&=u_i\,,\\
		\Delta_{m,1}&=\Delta_m-\bigg(N+1-2\left\lfloor\fft{N+1}{2}\right\rfloor\bigg)\,,\\
		n_{1,i}&=n_i\,,\\
		\mft_1&=\mft\,.
	\end{split}\label{V52:identification}
\end{equation}
To match the Bethe potentials, we have used the inversion formula (\ref{polylog:inversion}) under the assumptions
\begin{equation}
	0<\Re[u_{i}-u_{j}+\pi\Delta_I]<2\pi\,.
\end{equation}
To be more precise, the Bethe potential (\ref{V:V52}) matches the expression in \cite{Bobev:2023lkx} up to gauge holonomy independent terms. Such a difference does not affect the BAE, however, and therefore the numerical BAE solutions of \cite{Bobev:2023lkx} can be utilized in the generic $\mN=2$ conventions of this paper without modification.

\medskip

Next we provide numerical data that supports the all-order $1/N$ expansion of the $V^{5,2}$ Bethe potential given in (\ref{mW:Cardy:0:V52:saddle}). The list of $N_f$ and $\Delta$-configurations for which we confirmed (\ref{mW:Cardy:0:V52:saddle}) is given as follows ($\Delta=(\Delta_I,\Delta_q,\Delta_{\tq},\Delta_m)$). 

\medskip

\noindent\textbf{Case 1.} $\Delta=(\Delta_1,\fft43-\Delta_1,\fft23,\fft13,\fft13,0)$
\begin{equation}
	\begin{alignedat}{3}
		N_f&\in\{1,2,3,4,5\}&\quad&\&&\quad \Delta_1&=\fft23\,,\\
		N_f&\in\{1,2,3\}&\quad&\&&\quad \Delta_1&=\fft12\,,
	\end{alignedat}\label{V52:case1}
\end{equation}
\noindent\textbf{Case 2.} $\Delta=(\Delta_1,\fft43-\Delta_1,\fft23,\fft13,\fft13,\Delta_m)$
\begin{equation}
	\begin{alignedat}{3}
		N_f&\in\{1,2,3\}&\quad&\&&\quad (\Delta_1,\Delta_m)&\in\bigg\{(\fft59,\fft{N_f}{9}),(\fft{7}{12},\fft{N_f}{15})\bigg\}\,,
	\end{alignedat}\label{V52:case2}
\end{equation}
\noindent\textbf{Case 3.} $\Delta=(\Delta_1,\fft43-\Delta_1,\fft23,\Delta_q,\fft23-\Delta_q,\Delta_m)$
\begin{equation}
	\begin{alignedat}{3}
		N_f&\in\{1,2\}&\quad&\&&\quad (\Delta_1,\Delta_q,\Delta_m)&=(\fft23-\fft{1}{2\pi},\fft16,N_f(\fft23-\fft2\pi))\,.
	\end{alignedat}\label{V52:case3}
\end{equation}
For the above listed $N_f$ and $\Delta$-configurations, we estimate numerical coefficients
\begin{equation}
	\hat g_{3/2}^\text{(lmf)}(N_f,\Delta)\qquad\&\qquad\hat g_0^\text{(lmf)}(N_f,\Delta)
\end{equation}
together with the associated standard errors $\sigma_{3/2}$ \& $\sigma_0$ in the \texttt{LinearModelFit} for the $V^{5,2}$ Bethe potential, namely
\begin{equation}
	\fft{1}{2\pi}\Im\mV^{V^{5,2}}[u_\star;\Delta,n]=\hat g_{3/2}^\text{(lmf)}(N_f,\Delta)\hat N_{N_f,\Delta}^\fft32+\hat g_0^\text{(lmf)}(N_f,\Delta)\,,
\end{equation}
based on the numerical BAE solutions $\{u_\star\}$ constructed in \cite{Bobev:2023lkx} for $N=101\sim301~(\text{in steps of }10)$ at \texttt{WorkingPrecision} $200$. The leading order coefficient is then compared with the corresponding analytic expression in (\ref{mW:Cardy:0:V52:saddle}), namely
\begin{equation}
	\begin{split}
		\hat g_{3/2}(N_f,\Delta)=\fft{\pi\sqrt{N_f\tDelta_1\tDelta_2\tDelta_3\tDelta_4}}{3}\,,
	\end{split}
\end{equation}
by presenting the error ratio
\begin{equation}
	R_{3/2}(N_f,\Delta) = \fft{\hat g_{3/2}^\text{(lmf)}(N_f,\Delta)-\hat g_{3/2}(N_f,\Delta)}{\hat g_{3/2}(N_f,\Delta)}\,.
\end{equation}
The following tables summarize the numerical data described above.

\medskip

%%%%%
\noindent\textbf{Case 1. $\Delta_1=\fft23$}
%%%%%
%
\begin{center}
	\footnotesize
	\begin{tabular}{ |c||c|c|c|c| } 
		\hline
		& $R_{3/2}$ & $\sigma_{3/2}$ & $\hat g_0^\text{(lmf)}$ & $\sigma_0$\\
		\hline\hline
		$N_f=1$ & $-4.530{\times}10^{-31}$ & $1.278{\times}10^{-31}$ & $-0.13945567062297404931$ & $4.118{\times}10^{-28}$ \\
		\hline
		$N_f=2$ & $7.423{\times}10^{-25}$ & $2.745{\times}10^{-25}$ & $-0.18888453456219889345$ & $8.850{\times}10^{-22}$\\
		\hline
		$N_f=3$ & $1.260{\times}10^{-19}$ & $5.286{\times}10^{-20}$ & $-0.29657318363534945350$ & $1.705{\times}10^{-16}$\\ 
		\hline
		$N_f=4$ & $3.228{\times}10^{-17}$ & $1.473{\times}10^{-17}$ & $-0.45154281492584592878$ & $4.756{\times}10^{-14}$\\ 
		\hline
		$N_f=5$ & $1.335{\times}10^{-15}$ & $6.463{\times}10^{-16}$ & $-0.65205454918709992787$ & $2.089{\times}10^{-12}$\\ 
		\hline
	\end{tabular}
\end{center}
%

%%%%%
\noindent\textbf{Case 1. $\Delta_1=\fft12$}
%%%%%
%
\begin{center}
	\footnotesize
	\begin{tabular}{ |c||c|c|c|c| } 
		\hline
		& $R_{3/2}$ & $\sigma_{3/2}$ & $\hat g_0^\text{(lmf)}$ & $\sigma_0$\\
		\hline\hline
		$N_f=1$ & $-1.307{\times}10^{-25}$ & $3.377{\times}10^{-26}$ & $-0.14109222832291037605$ & $1.088{\times}10^{-22}$ \\
		\hline
		$N_f=2$ & $1.239{\times}10^{-20}$ & $4.070{\times}10^{-21}$ & $-0.20242609578739029964$ & $1.312{\times}10^{-17}$\\
		\hline
		$N_f=3$ & $3.814{\times}10^{-17}$ & $1.408{\times}10^{-17}$ & $-0.32947376221940705368$ & $4.545{\times}10^{-14}$\\ 
		\hline
	\end{tabular}
\end{center}
%

%%%%%
\noindent\textbf{Case 2. $(\Delta_1,\Delta_m)=(\fft59,\fft{N_f}{9})$}
%%%%%
%
\begin{center}
	\footnotesize
	\begin{tabular}{ |c||c|c|c|c| } 
		\hline
		& $R_{3/2}$ & $\sigma_{3/2}$ & $\hat g_0^\text{(lmf)}$ & $\sigma_0$\\
		\hline\hline
		$N_f=1$ & $-6.905{\times}10^{-25}$ & $1.761{\times}10^{-25}$ & $-0.14076080517355255410$ & $5.675{\times}10^{-22}$ \\
		\hline
		$N_f=2$ & $2.478{\times}10^{-19}$ & $7.910{\times}10^{-20}$ & $-0.19211160033287729944$ & $2.550{\times}10^{-16}$\\
		\hline
		$N_f=3$ & $3.292{\times}10^{-16}$ & $1.175{\times}10^{-16}$ & $-0.30398703304578607507$ & $3.793{\times}10^{-13}$\\ 
		\hline
	\end{tabular}
\end{center}
%

%%%%%
\noindent\textbf{Case 2. $(\Delta_1,\Delta_m)=(\fft{7}{12},\fft{N_f}{15})$}
%%%%%
%
\begin{center}
	\footnotesize
	\begin{tabular}{ |c||c|c|c|c| } 
		\hline
		& $R_{3/2}$ & $\sigma_{3/2}$ & $\hat g_0^\text{(lmf)}$ & $\sigma_0$\\
		\hline\hline
		$N_f=1$ & $-5.551{\times}10^{-27}$ & $1.480{\times}10^{-27}$ & $-0.14004350452924789065$ & $4.771{\times}10^{-24}$ \\
		\hline
		$N_f=2$ & $4.841{\times}10^{-21}$ & $1.639{\times}10^{-21}$ & $-0.19117259832590889123$ & $5.285{\times}10^{-18}$\\
		\hline
		$N_f=3$ & $1.685{\times}10^{-17}$ & $6.428{\times}10^{-18}$ & $-0.30199934416516391009$ & $2.074{\times}10^{-14}$\\ 
		\hline
	\end{tabular}
\end{center}
%

%%%%%
\noindent\textbf{Case 3. $(\Delta_1,\Delta_q,\Delta_m)=(\fft23-\fft{1}{2\pi},\fft16,N_f(\fft23-\fft2\pi)$}
%%%%%
%
\begin{center}
	\footnotesize
	\begin{tabular}{ |c||c|c|c|c| } 
		\hline
		& $R_{3/2}$ & $\sigma_{3/2}$ & $\hat g_0^\text{(lmf)}$ & $\sigma_0$\\
		\hline\hline
		$N_f=1$ & $-1.619{\times}10^{-25}$ & $4.173{\times}10^{-26}$ & $-0.14096914088723067753$ & $1.345{\times}10^{-22}$ \\
		\hline
		$N_f=2$ & $2.628{\times}10^{-20}$ & $8.540{\times}10^{-21}$ & $-0.20095533539635298960$ & $2.754{\times}10^{-17}$\\
		\hline
	\end{tabular}
\end{center}
%

%%%%%
\subsection{$Q^{1,1,1}$ theory}\label{app:Bethe:Q111}
%%%%%
The Bethe potential (\ref{V:Q111}) and the TTI (\ref{TTI:Q111}) of the $Q^{1,1,1}$ theory can be identified with the expressions presented in \cite{Bobev:2023lkx} after the substitutions
\begin{equation}
	\begin{split}
		(u_{1,i},u_{2,i})&=(u_i,\tu_i)\,,\\
		(\Delta_{m,1},\Delta_{m,2})&=(N+N_f-2\left\lfloor\fft{N_f}{2}\right\rfloor,-N-N_f+2\left\lfloor\fft{N_f}{2}\right\rfloor)\,,\\
		(n_{1,i},n_{2,i})&=(n_i+N,\tn_i+N)\,,\\
		(\mft_1,\mft_2)&=(0,0)\,.
	\end{split}\label{Q111:identification}
\end{equation}
To match the Bethe potentials, we have used the inversion formula (\ref{polylog:inversion}) under the assumptions
\begin{equation}
	0<\Re[u_{i}-\tu_{j}+\pi\Delta_{1,2}]<2\pi\qquad\&\qquad 0<\Re[\tu_{i}+\pi\Delta_{q_{1,2}}]<2\pi\,.
\end{equation}
As in the $V^{5,2}$ theory case, the Bethe potentials are matched up to gauge holonomy independent terms that do not affect the BAE.

\medskip

Next we present numerical data that supports the all-order $1/N$ expansion of the $Q^{1,1,1}$ Bethe potential given in (\ref{mW:Cardy:0:Q111:saddle}). The list of $N_f$ and $\Delta$-configurations satisfying the constraints (\ref{Q111:special}) for which we confirmed (\ref{mW:Cardy:0:Q111:saddle}) is given as follows: 
\begin{equation}
	\begin{alignedat}{3}
		N_f&\in\{1,2,3,4,5\}&\quad&\&&\quad\Delta_1&=\fft12\,,\\
		N_f&\in\{1,2,3\}&\quad&\&&\quad\Delta_1&\in\bigg\{\fft38,\fft{5}{12},\fft37\bigg\}\,.
	\end{alignedat}\label{Q111:case}
\end{equation}
For the above listed $N_f$ and $\Delta$-configurations, we estimate numerical coefficients
\begin{equation}
	\hat g_{3/2}^\text{(lmf)}(N_f,\Delta)\,,\qquad\&\qquad\hat g_0^\text{(lmf)}(N_f,\Delta)\,,
\end{equation}
together with the associated standard errors $\sigma_{3/2}$ \& $\sigma_0$ in the \texttt{LinearModelFit} for the $Q^{1,1,1}$ Bethe potential, namely
\begin{equation}
	\fft{1}{2\pi}\Im\mV^{Q^{1,1,1}}[u_\star;\Delta,n]\Big|_\text{(\ref{Q111:special})}=\hat g_{3/2}^\text{(lmf)}(N_f,\Delta)\hat N_{N_f}^\fft32+\hat g_0^\text{(lmf)}(N_f,\Delta)\,,
\end{equation}
based on the numerical BAE solutions $\{u_\star\}$ constructed in \cite{Bobev:2023lkx} for $N=101\sim301~(\text{in steps of }10)$ at \texttt{WorkingPrecision} $200$. The leading order coefficient is then compared with the corresponding analytic expression in (\ref{mW:Cardy:0:Q111:saddle}), namely
\begin{equation}
	\begin{split}
		\hat g_{3/2}(N_f,\Delta)=\fft{\pi\sqrt{N_f}}{3\sqrt{3}}\,,
	\end{split}
\end{equation}
by presenting the error ratio
\begin{equation}
	R_{3/2}(N_f,\Delta) = \fft{\hat g_{3/2}^\text{(lmf)}(N_f,\Delta)-\hat g_{3/2}(N_f,\Delta)}{\hat g_{3/2}(N_f,\Delta)}\,.
\end{equation}
The following table summarizes the numerical data we used for this analysis.

\medskip

\begin{center}
	\footnotesize
	\begin{tabular}{ |c||c|c|c|c| } 
		\hline
		& $R_{3/2}$ & $\sigma_{3/2}$ & $\hat g_0^\text{(lmf)}$ & $\sigma_0$\\
		\hline\hline
		$N_f=1$ & $-9.038{\times}10^{-20}$ & $2.846{\times}10^{-20}$ & $-0.12179382823357287453$ & $9.159{\times}10^{-17}$ \\
		\hline
		$N_f=2$ & $2.925{\times}10^{-15}$ & $1.114{\times}10^{-15}$ & $-0.060896914126385874431$ & $3.588{\times}10^{-12}$\\
		\hline
		$N_f=3$ & $8.415{\times}10^{-13}$ & $3.513{\times}10^{-13}$ & $0.018581373235204659187$ & $1.133{\times}10^{-9}$\\ 
		\hline
		$N_f=4$ & $2.296{\times}10^{-11}$ & $1.014{\times}10^{-11}$ & $0.12639484451282630333$ & $3.274{\times}10^{-8}$\\ 
		\hline
		$N_f=5$ & $2.206{\times}10^{-10}$ & $1.013{\times}10^{-10}$ & $0.26400260477995552485$ & $3.274{\times}10^{-7}$\\ 
		\hline
	\end{tabular}
\end{center}
The numerical estimates for the $Q^{1,1,1}$ Bethe potential do not depend on the $\Delta_1=\Delta_2$ value so the above table is valid for all cases listed in (\ref{Q111:case}).

%%%%%%%%%%%%%%%%%%%%%%%%%%%%
\bibliography{N=2TTISCI}
\bibliographystyle{JHEP}

%%%%%%%%%%%%%%%%%%%%%%%%%%%%

\end{document}